\documentclass{article}
\usepackage[T2A]{fontenc}
\usepackage[cp866]{inputenc} 
\usepackage[english]{babel}
\usepackage[tbtags]{amsmath}
\usepackage{amsfonts,amssymb,mathrsfs,amscd,comment}

\usepackage{amsmath}

\newcommand{\R}{\mathbb{R}}
\newcommand{\G}{{\Gamma}}
\newcommand{\BE}{{Boltzmann equation }}

\newcounter{proposition}

\begin{document}

\renewcommand{\contentsname}{\center{Contents}}

\renewcommand{\refname}{Bibliography}

\title{\bf\large\MakeUppercase{%
Boltzmann-type kinetic \\equations
 and discrete models}}
\author{A.\,V.~Bobylev}
\date{}

\maketitle

\makeatletter

\renewcommand{\@makefnmark}{}

\makeatother

\begin{abstract}
The known nonlinear kinetic equations (in particular, the wave kinetic equation
and the quantum Nordheim -- Uehling -- Uhlenbeck equations) are considered as a
natural generalization of the classical spatially homogeneous Boltzmann equation.
To this goal we introduce the general Boltzmann - type kinetic equation that depends 
on a function of four real variables $ F(x, y; v, w) $. The function $ F  $ is assumed to
satisfy certain commutation relations. The general properties of this equation
are studied. It is shown that the above mentioned kinetic equations correspond to
different forms of the function (polynomial) $ F $. Then the problem of discretization
of the general Boltzmann - type kinetic equation is considered on the basis of ideas which
are similar to those used for construction of discrete models of the Boltzmann
equation. The main attention is paid to discrete models of the wave kinetic
equation. It is shown that such models have a monotone functional similar
to Boltzmann $ H $-function. The existence and uniqueness theorem for global in time 
solution of the Cauchy problem for these models is proved. Moreover it is proved 
that the solution converges to the equilibrium solution when time goes to infinity.
The properties of the equilibrium solution and the connection with solutions of the 
wave kinetic equation are discussed. The problem of approximation of the Boltzmann-type 
equation by its discrete models and the problem of construction of normal discrete models 
are also discussed. The paper contains a concise introduction to the Boltzmann equation and its main properties. In principle, it allows to read the paper without any preliminary knowledge in 
kinetic theory.
\end{abstract}


{\tableofcontents}

  \section{Introduction}

      The classical Boltzmann equation, which is the main mathematical tool for description of
rarefied gases, occupies a very specific place in mathematical physics. Speaking about equations of
mathematical physics, we normally have in mind linear or non-linear equations in partial
derivatives. Nonlinear kinetic equations of the Boltzmann type belong to a bit different kind of
equations, though they contain partial derivatives together with multiple integrals. These equations
look too cumbersome at first glance. Perhaps, partly for this reason they are not included in standard
university courses for mathematicians. Nevertheless the interest of mathematicians all over the
world to this part of mathematical physics is growing fast in last decades.
It is very typical for history of physics and mathematics that some parts of physics, which
looked not very clear (even for physicists) after first fundamental discoveries, become with time
rather a part of mathematics. This can be a slow process. Of course, it does not mean that its
importance for physics disappeared. The famous examples are Newton's mechanics and Maxwell's
electrodynamics. Something like that happened and still is happening with kinetic theory of gases
founded by Maxwell \cite{Max}, Boltzmann \cite{Bol} and their predecessors approximately 150 years ago.  

The Boltzmann equation was firstly published in 1872 \cite{Bol}. This equation has a very
interesting history, we mention only one important point. Formally this equation was supposed to
describe more accurately, than the known at that time equations of gas dynamics, a behaviour of
rarefied gas of particles interacting by laws of classical mechanics. The first conclusion made by
Boltzmann in his paper cited above was his $  H$-theorem that formally proves the existence of
monotone decreasing it time functional on any solution of his equation. In fact, it was a discovery
of the mechanical meaning of entropy, the result which is probably more important that the
Boltzmann equation itself. 
The immediate consequence was that the solution of the Boltzmann equation
cannot be invertible in time in contrast with equations of Newtonian mechanics. 
This caused certain
doubts in this equation, especially among mathematicians. The famous remark by Zermelo \cite{Zer}
based on the Poincare recurrence theorem was made at the end of 19th century. This question and
others were clarified since then, the validity of the Boltzmann equation is justified rigorously at
least for short time-intervals.
We briefly describe below a progress in mathematical results for the Boltzmann equation. Only a
few such results were obtained before 1960s. These are, in particular, works by Hilbert \cite{Hil},
Carleman \cite{Car}, Morgenstern \cite{Morg}, and Grad \cite{Gra}. Of course, the great contribution of N.N.
Bogolyubov \cite{Bog} who proposed in 1946 his famous method of derivation of kinetic equations
from dynamics should me mentioned. Some of above cited works were done at the formal level of
mathematical rigour, but it is inevitable at the early stage of development of any mathematical theory related to
physics. The Chapman-Enskog method \cite{Cha}, \cite{Ens}, invented by physicists for transition to
hydrodynamics from the Boltzmann equation became a standard tool for study of dynamical systems with small parameter.

The number of rigorous mathematical results in kinetic theory began to grow faster in 1960-70s.
In particular, we mention: (1) the mathematically rigorous theory of the Cauchy problem for the
linearized Boltzmann equation constructed in works by Arsen'ev \cite{Ars} (for short-range
intermolecular potentials) and Ellis, Pinsky \cite{EP} (for power-like potentials with Grad's angular
cut-off); (2) the complete theory of existence and uniqueness of solutions to the Cauchy problem for
the spatially homogeneous Boltzmann equation for a wide class of potentials was developed by Arkeryd \cite{Ar1}; (3) the first global in
time existence and uniqueness theorems for the Cauchy problem for the spatially inhomogeneous
Boltzmann equation by Maslova and Firsov \cite{Mas} and Ukai \cite{Uka} under assumption that the initial
conditions are sufficiently close to equilibrium; (4) the proof by Lanford \cite{Lan} of the first validation
theorem for the Boltzmann equation in the Boltzmann-Grad limit for hard spheres on the time-
interval of order of the particle mean free path. For the sake of brevity we do not discuss the related
results obtained for other two famous classical nonlinear kinetic equations introduced respectively by
Landau \cite{Land} and Vlasov \cite{Vla}.
The kinetic theory and, in particular, the theory of the Boltzmann equation is gradually
becoming a more or less regular part of mathematical physics. Mathematical conferences on that
subject become more and more frequent in Europe, USA and Japan in last two decades of 20th
century. There was a sort of competition of pure mathematicians from different countries to prove a
global existence theorem for the Boltzmann equation with initial data far from equilibrium. Finally
the result was obtained by DiPerna and Lions in 1989 \cite{DPL}. The Fields Medal obtained by P.-L.
Lions apparently attracted a lot of good young mathematicians to kinetic theory, especially in
France. One of them, C. Villani, has obtained another Fields Medal in 2010 for his works in this
field of mathematical physics. He also wrote an excellent review \cite{Vil} of mathematical results on
the Boltzmann equation. This review with its renewed on-line version and the book \cite{CIP} contain
most of important mathematical results for the Boltzmann equation obtained before 2005.
A review of some more recent results can be found e.g. in the book \cite{Bobook}.
The development of kinetic theory in last decades is connected with applications of its ideas to
various unusual objects, which are far from traditional rarefied gases. It is related also to fast
development of computers and numerical methods. Kinetic equations are used now for modelling
of traffic flows, distribution of "active particles" \, (viruses, etc.) in biology, socio- economic
processes. Of course, many of these equations are perhaps "too young" \, to become a subject of
rigorous mathematical study. On the other hand, there are some classes of relatively "old" \, kinetic
equations, which are actively used by physicists since 1960th (see e.g. \cite{GK}, \cite{Zakh}). These are the
so-called "wave kinetic equations" \,used in the theory of weak turbulence. There are some
mathematical results on these equations (see e.g. \cite{EV}, \cite{DymK} and references therein), but still
many questions like a long time asymptotic behaviour of solutions remain unclear. One of goals of
the present paper is to study these and similar equations (for example, the quantum Nordheim--Uehling--Uhlenbeck equation \cite{Nor}, \cite{UU}) from a unified point of view as particular cases of the
general class of Boltzmann-type equations \cite{BK}. Our approach to this class of equations is, to some
extent, close to approach of L. Arkeryd in \cite{Ar4}. We apply to this class of equations some methods
used for the classical Boltzmann equations. In particular, the construction and properties of discrete
models of Boltzmann-type equations are discussed. The main attention is paid to the wave kinetic
equation (WKE), it is proved that the solutions of any normal discrete model of WKE tends to
equilibrium distribution when time tends to infinity. The consequences of this fact for solutions of
WKE are discussed in detail.
It should be noted that we do not discuss in this paper important question related to
validation of various Boltzmann-type equations, i.e. their derivation from some more general
mathematical model. It is not easy question. For example, it took more than a hundred years to get
a mathematical proof (Lanford's theorem \cite{Lan}) of connection between the classical Boltzmann
equation and the system of N hard spheres when N tends to infinity. A recent mathematical result on
derivation of WKE from non-linear Schr\"odinger equation (in the presence of random force field) in
\cite{DymK} looks very promising in that sense.

The paper is organized as follows. Section 2 presents a concise introduction to key ideas and
technique that lead from Hamiltonian mechanics to the classical Boltzmann equation. In principle, it
allows to read the paper without preliminary knowledge in kinetic theory. On the other hand,
important notations and some technical tools are introduced there. We begin with the $  N$-particle
system described by Newton equations in Hamiltonian form. Then we introduce the notion of $ N $-particle distribution function and the Liouville equation. Following the Grad version of BBGKY-hierarchy, we finally present a formal derivation of the Boltzmann equation for hard spheres. The
formal generalization of that equation to the case of any reasonable intermolecular potential is also
made. Then we study all general properties of the Boltzmann equation, including conservation laws
and the $  H $-theorem.

The general Boltzmann-type kinetic is introduced at the beginning of Section 3 as a natural
generalization of the classical Boltzmann equation from Section 2. This new equation depends on
an arbitrary function $  F(x_1, x_2, x_3, x_4)$ of four variables. For Boltzmann equation we have 
$  F =
x_3 x_4 - x_1 x_2 $. 
Other above mentioned kinetic equations correspond to different polynomial
forms of $ F $. We show that all these equations can be studied from a unified point of view. Different
forms of the general kinetic equation are constructed for the $ 3d $-case (Proposition \ref{Prop3.1}). The
generalization to $  d $-dimensional case with 
$ d \leq 2 $ is also considered. The weak form of
the general kinetic equation (the equation for average values) and conservation laws are also
discussed. We define a class of functions F that can lead to an analogue of Boltzmann's $  H $-theorem
(Proposition \ref{Prop3.2}). Then we introduce discrete kinetic models of the general Boltzmann-type kinetic
equation by using the formal analogy with discrete velocity models of the Boltzmann equation.
Similarly we define a notion of a normal discrete model. Then we prove Theorem \ref{Th3.1} on main
properties of normal discrete models which posses an analogue of $ H $-theorem.

Section 4 is devoted to some properties of solutions to normal discrete kinetic models of
WKE. The main result of this section is the proof of convergence of any positive solution of this
model to unique equilibrium solution. This result is formulated at the beginning of Section 4 (Theorem
4.1). The proof is given in the rest of that section. First we construct the solution for any positive
initial data and prove its global in time existence and uniqueness (Lemma \ref{lem4.1}). Then we construct a
positive stationary solution of the model and prove its uniqueness under given invariants (mass and
energy) in Lemma \ref{lem4.2}. Then we improve some estimates for strictly positive initial data and
complete the proof of convergence to equilibrium by more or less standard methods of the theory of
ODEs.

Appendixes A and B contain some facts based on known properties of discrete kinetic models.   These results  were mentioned  in Sections 3--4.  Appendix A explains how to construct the normal discrete models introduced in Section 3. It is important because the main result of Section 4 (Theorem 4.1) is valid only for this particular class of models. Appendix B explains how to approximate the Boltzmann-type kinetic equation by a sequence of discrete models when the order of the model, i.e. the number of its discrete points, tends to infinity. It is interesting that the proof of approximation is based on deep results of the theory of numbers.


\section{From particle dynamics to the Boltzmann equation}

\numberwithin{equation} {section}


\subsection{$N$-particle dynamics and
modeling of rarefied gases}

We consider $  N \ge 1$ identical particles with mass $  m=1  $.
This system is characterized  by  a $ 6N-   $dimensional
phase vector  $ Z_N=\{z_1,...,z_N\}$ with components
 $ z_i=(x_i,v_i)   $, where $  x_i \in \mathbb{R}^3   $
 and $  v_i \in \mathbb{R}^3   $ denote respectively
a position  and a velocity of  $ i^{th}   $ particle,
$ i=1,...,N   $.
Usually it is assumed below that the particles interact
via given pair potential $ \Phi(r)   $, where $ r>0   $
 stands for the distance between two interacting particles.
We also assume that $  \Phi(r) \to 0 $ if $ r \to\infty   $.
 The equations of motion of the
system have the following Hamiltonian form (see any
textbook in classical mechanics, e.g.~\cite{LL1}):
\begin{equation}\label{1.1}
\begin{split}
  &
  \partial_t x_i= \partial H_N /\partial v_i,
   \quad \partial_t v_i = - \partial H_N /\partial x_i,\\
 &
  H_N = \frac{1}{2} \sum \limits_{i=1}^N \, |v_i|^2
  + \sum \limits_{1\leq i < j\leq N} \Phi(|x_i-x_j|),
  \\
  &
 x_i(0)=x_i^{(0)},\; v_i(0)=v_i^{(0)},
  \quad i,j=1,...,N.
\end{split}
\end{equation}
Thus the temporal evolution of the system can be
understood as the motion of the phase point
$ Z_N=\{z_1,...,z_N\}$ in the phase space
 $   \mathbb{R}^{6N} $. This motion obeys
 conservation laws of energy  $ E_N \in \mathbb{R}$ and momentum
  $ P_N\in \mathbb{R}^3$ respectively
 \begin{equation} \label{1.2}
 E_N=H_N[Z_N(t)]=\mathrm{const.},\quad
 P_N=
 \sum \limits_{i=1}^N \,v_i(t)= \mathrm{const.}
 \end{equation}
 The laws~\eqref{1.2} follow directly from Eqs.~\eqref{1.1}.

 In principle, the above described $ N $-particle system
  can be used for modeling of real gases or liquids.
 Then the main problem is that the number of particles
 $ N $ is of order of $ 10^{23}$ ( Avogadro's number). 
 It is intuitively clear that the case of rarefied gas 
 is easier for description then the case of dense gas.
 Indeed in the low density limit we get a free
 molecular flow, i.~e. each particle moves
 independently with constant velocity. Consequently our main
 assumption will be the following: a typical distance
 between particles is much greater then effective
 diameter $ d $ of the potential. It is known that
the typical ''size'' $ d $ of a molecule of the air
is roughly equal to $ d \approx 3.7 \cdot 10 ^{-8} cm $,
whereas the
number density $ \bar{n} $ of the air is about
$ {\bar{n}} \approx 2.5 \cdot 10 ^{19} cm^{-3} $ under 
normal conditions. The
 inequality
\begin{equation}\label{1.3}
\delta = \bar{n}d^3 << 1
\end{equation}
is the well-known criterion for ideal gas. Note
that $\delta \approx 10 ^{-3} $ for air under normal
conditions on the surface of the Earth.
This parameter $\delta $ is decreasing with height.
Therefore kinetic equations (in particular,
the classical Boltzmann equation for rarefied gases)
are important
for applications in the space science and technology.

The simplest model intermolecular potential
$ \Phi(r) $ corresponds to particles interacting like
 hard spheres of diameter $ d $. Then we formally obtain
\begin{equation}\label{1.4}
\Phi_{HS}(r)=
\left\{\begin{aligned}
&\infty \quad  \text {if} \quad & 0< r \leq d\\
&0   \quad  &  \text{otherwise.}
\end{aligned}\right.
\end{equation}
Another well-known model corresponds to power-like
repulsive potentials
\begin{equation}\label{1.5} 
\Phi(r)=\frac\alpha{r^n},\quad
\alpha > 0,\; n\geq 1,
\end{equation}
including the Coulomb case $n=1$ with any sign of
$\alpha$.

 In the next section we discuss some probabilistic
 aspects of kinetic theory of gases.


\numberwithin{equation} {section}

\subsection{Distribution functions and Liouville equation}

We introduce an important notion of one-particle
distribution function $ f(x,v,t)$ (the words
''one-particle'' are usually omitted below for the
sake of brevity). The physical meaning of this
function is the following: the average number of
particles in any measurable set $ \Delta \in
 \mathbb{R}^3 \times  \mathbb{R}^3 $ is given by
 equality
 \begin{equation}\label{2.1}
n_\Delta (t)=\int\limits_\Delta \,dx dv f(x,v,t).
\end{equation}
In other words, $ f(x,v,t) $ is the density of
number of particles in the phase space.
Usually we assume that the initial data
\begin{equation}\label{2.2} 
f(x,v,0)=f^{(0)}(x,v)
\end{equation}
are given.
How to find the distribution function $ f(x,v,t) $ for $  t>0   $?
This is, in a sense, the main problem of kinetic
theory.
It can be shown that for some special physical systems,
like rarefied gases, the temporal evolution of
distribution function $ f(x,v,t) $  is described by 
so-called "kinetic equation"
\begin{equation}\label{2.3} 
f_t= A(f),
\end{equation}
where $  A(f) $ is a nonlinear operator acting on
$   f   $. We usually assume that the initial value
problem~\eqref{2.2}--\eqref{2.3} has a unique solution
$  f(x,v,t) $ on some time interval $ 0 \leq t \leq T$.

Let us consider the simplest kinetic equation connected
with the system~\eqref{1.1}. Omitting index $ i=1$ we
obtain equations of free motion
\begin{equation}\label{2.4}
x_t=v,   \quad     v_t=0.
\end{equation}
The motion of one particle can also be described
by  the distribution function $ f(x,v,t) $, having
in that case the meaning of probability density if
\begin{equation}\label{2.5} 
\int \limits_{\mathbb{R}^3\times\mathbb{R}^3}\!
dx dv
f^{(0)}(x,v)=1
\end{equation}
in the notation of~\eqref{2.2}. The solution of
Eqs.~\eqref{2.4} is obvious:
\[
x(t)=x(0) + v(0) t, \quad v(t)=v(0).
\]
Therefore $f(x,v,t)$, satisfying conditions~\eqref{2.2}, reads
\begin{equation}\label{2.6}  
f(x,v,t)=f^{(0)}(x-vt, v) = \exp \left(
-tv\cdot \partial _x \right)\, f^{(0)}(x,v)\,.
\end{equation}
Here and below dot denotes the scalar product
in $\mathbb{R}^3$.
We can check by differentiation that
\begin{equation}\label{2.7}
f_t + v\cdot \partial_x f = 0.
\end{equation}
This is the simplest kinetic equation. Note  that
 kinetic equation~\eqref{2.7} is exactly equivalent
  to dynamical equation~\eqref{2.4}.
The probabilistic description is caused only
by uncertainty in initial conditions.

Let us now extend these arguments to the case
of $ N  $ non-interacting particles. We
consider Eqs.~\eqref{1.1} with $\Phi(r)\equiv 0$
and obtain
\begin{equation}\label{2.8}
\begin{split}
\partial_t x_i=v_i, \quad  \partial_t v_i=0 ;\\
x_i(0)= x_i^{(0)},\quad v_i(0)=v_i^{(0)},
\qquad i=1,\dots, N.
\end{split}
\end{equation}
Thus we have $ N $ independent vector equations
for each particle. It is natural to introduce
 $ N $~-particle distribution function
 $ F_N(z_1,\dots,z_N;\,t)$,\; $z_i=(x_i,v_i)$, \;
 \; $1 \leq i \leq N $, with a meaning of a probability
  density in the  $ N $~-particle phase space $ \mathbb{R}^{6N}$.
   The initial condition reads
\begin{equation}\label{2.9} 
\begin{split}
F_N |_{t=0}= F_N^{(0)}(z_1, \dots, z_N),\\
\int \limits_{\mathbb{R}^{6}\times \dots \times \mathbb{R}^{6}}
\! \! \! \!dz_1\dots dz_n \;F_N^{(0)}(z_1, \dots, z_N) =1.
\end{split}
\end{equation}

\noindent
\textbf{Remark.} \;{
Here and below we use notations like
$ F_N(z_1,\dots,z_N) $ (with capital $F$) for
various multi-particle distribution functions,
which have a meaning of probability density. These
functions are always normalized by one in the
whole phase space. The notations like
$ f_N(z_1,\dots,z_N)$ will be used for slightly
different class of functions related to
equality~\eqref{2.1}. The difference disappears
for trivial case $N=1$.}

Then it is easy to see that
\begin{multline*}
F_N(z_1, \dots, z_N; t)= F_N^{(0)}[z_1(t), \dots, z_N(t)],
\\
z_i(t)= (x_i-v_it, v_i), \quad
\;i=1, \dots, N.
\end{multline*}
Note that
 \begin{equation} \label{2.10}
\left(\partial_t + \sum \limits_{i=1}^N \,
v\cdot \partial_{x_i}\right)
F_N(x_i,v_i, \dots,x_N,v_N; t)=0.
\end{equation}
Thus we obtain the simplest version of the Liouville
equation.

We assume that all particles are
identical and independently distributed at $t=0$,
i.e.
 \begin{equation} \label{2.11}   
F_N|_{t=0}=\prod \limits_{i=1}^N f^{(0)}(x_i,v_i).
\end{equation}
Then the similar factorization  holds for all $t>0$
 \begin{equation} \label{2.12}
F_N(z_1, \dots, z_N; t)=
\prod \limits_{i=1}^N f(z_i,t)=
\prod \limits_{i=1}^N f^{(0)}(x_i-v_i t, v_i).
\end{equation}
This property is known as ''the propagation of
chaos'' \cite{Kac}. It is self-evident for
non-interacting particles, but it also can be
proved as asymptotic property of more complex
multi-particle systems.

What changes if we consider the Hamiltonian
system~\eqref{1.1} with the nonzero potential
 $\Phi(r)\neq 0$?
Then we still can use the $N$~-particle
distribution function $F_N=\{z_1,\dots,z_N;t)$.
We shall see below that the equation
for $ F_N $ reads
 \begin{equation} \label{2.13} 
 \begin{split}
&
\left[
\partial_t + \sum \limits_{i=1}^N \,
\left(
v_i\cdot\partial_{x_i} -
\frac{\partial \Phi_N}{\partial x_i}\cdot  \partial_{v_i}
\right)
\right]
 F_N(x_1,v_1,\dots,x_N,v_N;\,t)=0\,,
\\
&
\Phi_N= \sum \limits_{1 \leq i <j \leq N}\,
 \Phi (|x_i-x_j|)\,;
 \quad i=1,\dots, N.
\end{split}
\end{equation}
This is the famous Liouville equation~\cite{LL1}.
It can be derived very easily. For a moment we
simplify our notations in the following way

\[
\begin{split}
z=\{z_1,\dots,z_N\} \in  \mathbb{R}^{6N},
\; z_i=(x_i,v_i);\\
w(z)=\{w_1,\dots,w_N\}
\in \mathbb{R}^{6N},\;
w_i=\left(      \frac{\partial H}{\partial v_i},
\;- \frac{\partial H}{\partial x_i}\right),\\
H=H_N=\frac1{2}\sum \limits_{i=1}^N \,|v_i|^2
+
\sum \limits_{1 \leq  i <j \leq N} \,\Phi(|x_i-x_j|);\\
F(z;t)=F_N(z_1,\dots,z_N;\,t).
\end{split}
\]
Then we can treat $ F(z,t)$ as a density of a
fluid in $ \mathbb{R}^{6N} $, which moves in
accordance with dynamical system
 \begin{equation} \label{2.14}
z_t=w(z)
\end{equation}
where $ w(z) $ is assumed to be a ''nice''  function.

Then the density $F(z;t)$ satisfies the continuity
equation
\begin{equation}\label{2.15}
F_t + \mathrm{div}_z F\,w =0,
\end{equation}
where $ \mathrm{div}_z $ denotes the divergence with
respect to $ z $. Simple calculation yields
\[\mathrm{div}_z \,w(z)= \sum \limits_{i=1}^N
\left(
\frac\partial {\partial x_i}\cdot \frac{\partial H}
{\partial v_i} - \frac\partial {\partial v_i}
\cdot \frac{\partial H}{\partial x_i}\right)=0.
\]
Hence, we obtain
\begin{equation}\label{2.16} 
F_t  + \sum\limits_{1}^{N}
\left(
\frac{\partial H} {\partial v_i}\cdot \frac{\partial F}
{\partial x_i} - \frac{\partial H}{\partial x_i}
\cdot \frac{\partial F}{\partial v_i} \right)
=0,
\end{equation}
i.e. the Liouville equation~\eqref{2.13} in slightly
different notations. The sum in~\eqref{2.16} is usually
called the Poisson brackets $ \{F,H\} $ (only for
particles with unit mass $ m=1 $)~\cite{LL1}.

The Liouville equation is very important because it
allows (at least formally) to build a bridge between
$ N $~-particle dynamics, kinetic theory and hydrodynamics.


\numberwithin{equation} {section}
\subsection{BBGKY-hierarchy}

It was assumed in Section 2.2 that $f_N(z_1,\dots,z_N;t)$
is integrable over the whole phase space ${\mathbb{R}}^{6N}$.
This assumption is sometimes too strong. For example it does
not allow to consider translationally invariant (in physical
space $ {\mathbb{R}}^3 $ ) systems with initial data like
$ f_2^{(0)}(x_1-x_2; v_1, v_2) $. Therefore it is more
convenient to consider $N-$~particle system confined in
a bounded domain $\Omega \subset {\mathbb{R}^3}$, say a box or
a sphere with ''large''  diameter $L$. The volume of $\Omega$
is denoted by $|\Omega|$. Then the phase state of $i^{\mathrm{th}}$
particle is described by the point
\begin{equation}\label{3.1}
z_i=(x_i,v_i) \in \Omega \times {\mathbb{R}^3} ,
\quad i=1, \dots,N.
\end{equation}

We can assume for simplicity that the walls of $ \Omega $
are specularly reflecting. Another possibility is to assume
that $ \Omega  $ is a periodic box. What is important is to keep
unchanged the total number $  N $ of particles inside $  \Omega $.

For the sake of brevity we will use below symbolic notations
\begin{equation}\label{3.2} 
\begin{split}
&
F_N(x_1,v_1;\dots;x_N,v_N;t)= F_N(1,2,\dots,N;t)\,, \\
&
\int\limits_
\Omega dx_i\int\limits_{\mathbb{R}^3}
dv_i \dots=
\int\limits_G di\dots\,; \quad
  G=\Omega\times {\mathbb{R}^3},
\;i=1,\dots,N,
\end{split}
\end{equation}
where $ F_N  $ is normalized by equality
\begin{equation}\label{3.3}
\int\limits_{G_N} d1 d2 \dots dN F_N(1,\dots,N;t) = 1,
\quad G_N=G^N ,
\end{equation}
and satisfies the Liouville equation~\eqref{2.3}
\begin{equation}\label{3.4} 
\begin{split}
&
\left( \partial_t + \sum \limits_{i=1}^N A_i -
\sum \limits_{1 \leq i< j \leq N} B_{ij}\right)
F_N=0\,,\quad A_i=v_i\cdot\partial_{x_i},
\\
&
B_{ij}=
\frac{\partial \Phi(|x_i-x_j|) }{ \partial  x_i}
\cdot ( \partial_{ v_i} -  \partial_{ v_j});
\quad i,j=1,\dots,N;\; i \neq j.
\end{split}
\end{equation}
The function $   F_N(1,\dots,N;t) $ is assumed to be invariant
under permutations of its arguments $ (1,\dots,N)  $ because all $N $
particles are identical. This property is preserved
by the Liouville equation if it is valid at $t=0$.

We introduce $k$~-particle probability
 distributions by equalities
\begin{equation}\label{3.5}
\begin{split}
F_k^{(N)}(1,2,\dots,k)= \!\!\int\limits_{G_{N-k}} \!\!d(k+1)\dots dN\;
&
F_N(1,2,\dots,N),\\
&
 1\leq k \leq N-1.
 \end{split}
\end{equation}
Here and below the argument $ t $ is omitted in such cases
 when it does not cause any confusion. The next step is to
obtain a set of evolution equations for $  F_k^{(N)}(1,2,\dots,k;t) $.
We take any $ 1\leq k \leq N-1 $   and integrate
Eq.~\eqref{3.4} over $ d(k+1)\dots dN $. The result reads
\begin{equation}\label{3.6} 
\left(\partial_t + L_k \right) F_k^{(N)}=
\Gamma_k^{(N)},\quad
L_k =\sum\limits_{i=1}^k \,A_i -
 \!\sum\limits_{1 \leq i < j \leq k}\!\! B_{ij},
\end{equation}
\begin{equation}\label{3.7}
\begin{split}
\Gamma_k^{(N)}= \int\limits_{G_{N-k}} \!\!
d(k+1)\dots dN
&
\left[
-\sum\limits_{i=k+1}^N A_i
+
 \sum\limits_{i=1}^k \sum\limits_{j=k+1}^N B_{ij}
+\right.\\
&\left.+\sum\limits_{i=k+1}^{N-1} \sum\limits_{j=i+1}^N B_{ij}
\right]\,
F_N(1,\dots,N).
\end{split}
\end{equation}
We assume that the boundary conditions guarantee
 that for any $1 \leq i \leq N  $
 \begin{equation}\label{3.8} 
\int\limits_{G}
di (v_i \cdot \partial_{x_i})\;
F_N(1,\dots,N; t)=0.
\end{equation}
Then  the first sum in~\eqref{3.7} disappears.
The third sum in~\eqref{3.7} also disappears
under natural assumption that
\begin{equation}\label{3.9}
F_N(1,\dots,N; \,t) \to 0 \quad \text{if}
\quad |v_i| \to \infty, \quad \text{for some}
\quad 1 \leq i \leq N.
\end{equation}
Then we obtain

\[
\Gamma_k^{(N)}= \sum\limits_{i=1}^k
\Gamma_{ik}^{(N)}, \quad
\Gamma_{ik}^{(N)}=
\sum\limits_{j=k+1}^N \;\int\limits_{G_{N-k}}\!\!
d(k+1)\dots dN \;B_{ij} \,F_N(1,\dots,N).
\]
Let us consider the first term of the sum for
$\Gamma_{ik}^{(N)}$. Then $j=k+1$ and we obtain
\[
\begin{split}
&
\int\limits_{G} \!d(k+1) B_{i\,k+1}
\!\!\int\limits_{G_{N-k-1}} \!\!d(k+2) \dots dN
F_N(1, \dots, N)=\\
&
=\int\limits_{G} d(k+1)\, B_{i\,k+1}\,
F_{k+1}^{(N) }(1, \dots, k+1),
\quad 1\leq k \leq N-1,
\quad F_N^{(N)}= F_N.
\end{split}
\]

If we take another value of $j$  in the sum
for $ \Gamma_{i\,k}^{(N)}  $, then the result will be the same.
It follows from symmetry of $ F_N(1, \dots, N)  $ with respect
to permutations. Moreover the operator $ B_{i\,k+1}  $
from~\eqref{3.4} can be replaced by
$$
\tilde{B}_{i\,k+1}  =\frac{\partial
\Phi(|x_i-x_{k+1}|)}{\partial x_i}\cdot
 \frac{\partial}{\partial v_i}
 $$
because of conditions~\eqref{3.9}. Therefore
we obtain
\begin{equation*}
\Gamma_k^{(N)}=(N-k)\sum\limits_{i=1}^k \,
\int\limits_{G} \! d(k+1) \tilde{B}_{i\,k+1}\,
F_{k+1}^{(N)}(1, \dots, k+1).
\end{equation*}
We substitute this formula into Eqs.~\eqref{3.6}
and get the following set of equations:
\begin{equation}\label{3.10}  
(\partial_t + L_k) F_k^{(N)} =(N-k) C_{k+1}
F_{k+1}^{(N)},\quad 1 \leq k \leq N-1,
\end{equation}
where $F_k^{(N)}(1, \dots,k) $ are given in Eqs.~\eqref{3.5},
\begin{equation}\label{3.11}
\begin{split}
&
F_N^{(N)}(1, \dots, N)=F_N(1, \dots, N),\\
&
L_k= \sum\limits_{i=1}^k \, v_i  \cdot \partial_{ x_i} -
 \sum\limits_{1 \leq i < j \leq k}\,
 \frac{\Phi(|x_i-x_{j}|)}{\partial x_i}\cdot
 {(\partial_{v_i} - \partial_{ v_j})},\\
 &C_{k+1} =\sum\limits_{i=1}^k \,C_{i\,k+1}\,, \\
 &
 C_{i\,k+1}\,F_{k+1}^{(N)}= \int\limits_{G} \! d(k+1)
 \frac{\Phi(|x_i-x_{k+1}|)}{\partial x_i}\cdot
 \partial_{v_i} F_{k+1}^{(N)} (1, \dots, k+1).
\end{split}
\end{equation}

 The set of equations~\eqref{3.10}--\eqref{3.11}
 is called (if we ignore its trivial modifications)
 the BBGKY-hierarchy. The BBGKY stands for
 Bogolyubov, Born, Green, Kirkwood and Yvon,
 the names of physicists who introduced
 independently this  system of equations (see
 e.g.~\cite{Bal,Bog}). The equation~\eqref{3.10} with $k=N $ also makes
 sense if we set $ F_N^{(N)}=F_N $, \,
 $F_N^{(N+1)} = 0 $. Then it will be the Liouville
 equation for $F_N $.
 Note that
 equations~\eqref{3.10} (the BBGKY-hierarchy
 for $N-$~particle dynamical system~\eqref{1.1} in the
 box $\Omega$) can be formally considered as
 exact equations. The only relevant assumption
 is equality~\eqref{3.8} related to interactions
 with boundaries of the box $\Omega$. This equality
 is fulfilled, in particular, for periodic box
 or the box with specularly reflecting walls.

 The BBGKY-hierarchy is
 a starting point for all classical kinetic
 equations. For example, let us assume that
 the interaction between particles is weak
 and replace $ \Phi(r) $ by  $ \varepsilon
 \bar{\Phi}(r) $ in Eqs.~\eqref{3.10} in the
 following form:
 \[
 \left(
 \partial_t + \sum \limits_{i=1}^N \, A_i-
 \varepsilon \!\!
 \sum \limits_{1 \leq  <j \leq N} \!\!B_{i j}
 \right)
  F_{k+1}^{(N)} =  \varepsilon \,(N-k)
  C_{k+1} F_{k+1}^{(N)}\,,\quad
  1 \leq k \leq N-1,
 \]
 in the notation of Eqs.~\eqref{3.4}, \eqref{3.11}.
 Then it is natural to consider the limit
 $$
 N \to \infty,\quad \varepsilon \to 0,
  \quad
 N\varepsilon =\mathrm{const.}
 $$
and to assume that
 $  F_{k}^{(N)} \to   F_k $ in that limit for all $ k=1,2, \dots$.
 Then we formally obtain the following infinite set
 of limiting equations for  $ F_{k} $:
 \[
 \begin{split}
& \left(
 \partial_t + \sum \limits_{i=1}^k \,
 v_i\cdot \partial_{x_i}
 \right)
  F_{k}(z_1,\dots,z_k;t)=
  \\
  &
  =(N\varepsilon) \sum \limits_{i=1}^k \,
  \int \limits_G dz_{k+1}
  \frac{\Phi(|x_i-x_{j}|)} {\partial x_i}\cdot
 \partial_{v_i} F_{k+1}(z_1,\dots,z_{k+1};t);
 \\
 &
 z_i=(x_i,v_i) \in G,\quad
 dz_i =dx_i dv_i,\quad i=1, \dots, k+1; \quad
 k=1, \dots
 \end{split}
 \]

  It is easy to verify that these equations admit
  a class of solutions in factorized form
  $$
  F_k(z_1,\dots,z_k;t) =
  \prod\limits_{i=1}^k \, F(z_i,t),
  \quad k =1, \dots,
  $$
  where $ F(x,v,t)$ satisfies the Vlasov
  equation~\cite{Vla}
  \[
  F_t + v \cdot F_x -
    (N\varepsilon)  F_v \cdot  \partial_x\,\int \limits_\Omega
    \! dy \,dw \,\Phi(|x-y|)\,F(y,w,t) = 0.
  \]
  The most important applications of the Vlasov
  equations are related to the Coulomb potential
  $\Phi(r)= \alpha/r $ (gravitational or
  electrostatic forces).

 On the other hand, in important case
 of potentials $ Phi(r) $ with compact support
 (strong interaction at short distances)
 it is
 more convenient to use a different
 approach. This approach will be considered 
 in the next section.

 %



\subsection{Hard spheres and Boltzmann-Grad limit}

We begin with the case of $N$ hard spheres of
of diameter $d$. Then the Liouville
 equation~\eqref{2.13} cannot be used
  directly because the potential
   $\Phi_{HS}(r)$~\eqref{1.4} is too singular. The
function $ F_N(x_1,v_1;\dots; x_N,v_N;t) $ is
defined in this case by the free flow equation

\begin{equation}\label{6.1} 
\partial_t F_N +  \sum\limits_{i=1}^N
v_i \cdot \partial_{x_i} F_N =0,
\end{equation}
valid in the domain

\begin{equation}\label{6.2}
B_N=\{x_i \in  \mathbb{R}^3:
\;|x_i -x_j| > d; \; i,j=1,\dots,N, \;
i\neq j\}.
\end{equation}
We add to this equation the boundary conditions
on each of $N(N-1)/2 $ boundary surfaces
$|x_i - x_j|=d \; (i,j=1, \dots, N; i\neq j)$ in
$\mathbb{R}^{3N}$. Taking, for example, $i=1,
j=2 $, we obtain, assuming the specular reflection
law,

\begin{equation}\label{6.3}
F_N(x_1,v_1;x_2,v_2;\dots;t)|_{|x_1-x_2|=d,\,
x\cdot u>0} =
F_N(x_1,v_1^\prime;x_2,v_2^\prime;\dots;t),
\end{equation}
where

\begin{equation}\label{6.4} 
\begin{split}
&
x=x_1-x_2= d\,n, \; n \in S^2, \quad u=v_1-v_2,\\
&
v_1^\prime=v_1 - n (u \cdot n),\quad
v_2^\prime=v_2 +n (u \cdot n).
\end{split}
\end{equation}
The surfaces that correspond to multiple collisions
of  $k \geq 3 $ particles are described by at least
$(k-1) \geq 2 $ equalities. For example, for
$ k =3 $ we need to satisfy simultaneously two
conditions like $|x_1 - x_2| = d $ and $|x_2-
x_3| =d$. These surfaces have a zero measure
 as compared with the case $k=2$ of pair collisions.
 Therefore we ignore multiple collisions.

 For brevity we assume that

 \begin{equation}\label{6.5}
 \begin{split}
 F_N \in L(G_N), \, G_N=G^N,\,G=\mathbb{R}^3\times
 \mathbb{R}^3,
 \\
 F_N(1,\dots,N;t)  = 0
 \quad \text{if} \quad |x_i - x_j| < d
 \end{split}
 \end{equation}
for at least one pair of indices $1 \leq i<j \leq N$.
Here and below we use symbolic notations~\eqref{3.2} from
Section 2.3, when it does not cause any confusion.
We also assume that $F_N \geq 0 $ is the probability
density in $G_N $ with usual normalization
condition~\eqref{3.3}. Our aim is to construct the
equation for one-particle probability density
$F_1^{(N)}(1)$ given in~\eqref{3.5} with $k=1$.
Note that this function can be equally defined by
equality
 \begin{equation}\label{6.6} 
  \begin{split}
F_1^{(N)}(1)=\!\! \int \limits_{G_{N-1}}\!\!
 d2 \dots dN \,\Psi
(1|2, \dots, N)\, F_N(1, \dots, N),
\\
 \Psi(1|2, \dots,N) =\prod\limits_{k=2}^N \eta
[d^2 - |x_1 - x_k|^2],
\end{split}
 \end{equation}
where $\eta(y)$ is the unit function

  \begin{equation}\label{6.7}
\eta(y)=
\begin{cases}
1, & \text{if $y>0$;} \\
0, & \text{otherwise.}
\end{cases}
  \end{equation}
We multiply Eq.~\eqref{6.1} by $ \Psi(1|2, \dots,N)$
and integrate over $ G_{N-1} $. The result reads

  \begin{equation}\label{6.8}
  \partial_t F_1^{(N)} + I_1 +
  \sum\limits_{k=2}^N I_k =0,
 \end{equation}
where

  \begin{equation}\label{6.9}
I_k =\!\! \int \limits_{G_{N-1}} \!\!d2 \dots dN\,
\Psi(1|2, \dots, N) \;A_k \;F_N(1,\dots, N),
\quad A_k = v_k \cdot \partial_{x_k}.
 \end{equation}
 We separated the term with $ k=1 $ in~\eqref{6.8}
 because all other terms in the sum are equal to
 $I_2$. Indeed, we always assume that all particles
 are identical and therefore $F_N(1, \dots, N )$
 is symmetric with respect to permutations of
 arguments $(2, \dots,N)$. Hence, $I_k=I_2 $
 for any $3 \leq k \leq N $ and therefore

 \begin{equation}\label{6.10}
 \sum\limits_{k=2}^N I_k = (N-1) I_2.
 \end{equation}
 The integral $I_2 $ can be written as

 \begin{equation}\label{6.11}
 I_2 = \!\! \int \limits_{|x_1-x_2|>d} \!\!\!\!dx_2\,
 \textrm{div}_{x_2} \int\limits_{ \mathbb{R}^3 }
 dv_2 \,v_2 \,\tilde{F}_2^{(N)}(x_1,v_1;x_2,v_2)\,,
 \end{equation}
 where
\begin{equation}\label{6.12}       
  \begin{split}
 \tilde{F}_2^{(N)}(x_1,v_1;x_2,v_2)=
   \tilde{F}_2^{(N)}(1,2)=\\
   \!\!
  = \int \limits_{G_{N-2}} \!\!d3 \dots dN
 F_N(1,,2, \dots,N)\,\prod\limits_{k=3}^N
 \eta[d^2 - |x_1 - x_k|^2].
 \end{split}
 \end{equation}
 We use the notation $  \tilde{F}_2^{(N)} $
 because  formally this function coincides with
  ${F}_2^{(N)}$ from  \eqref{3.5} only for $d=0$.
We apply the Gauss theorem to integral~\eqref{6.8}
and obtain after simple transformations
\[
\begin{split}
I_2= -\int\limits_{|y|>d}dy\, \mathrm{div}_y
\int\limits_{\mathbb{R}^3 } dv_2\, v_2
\tilde{F}_2^{(N)}(x_1,v_1; \, x_1-y, v_2)=\\
=
d^2\,\int\limits_{\mathbb{R}^3 \times S^2}
dv_2\, dn\, (v_2 \cdot n)
\tilde{F}_2^{(N)}(x_1,v_1; \,x_1-dn, v_2),
\end{split}
\]
where $n$ denotes the outward unit normal vector
to the unit sphere $S^2 $. It remains to evaluate
the integral $I_1$ in  \eqref{6.9}. We note that
$$
\Psi(1|2, \dots,N) A_1\,F_N(1, \dots,N)=
A_1 \Psi F_N  - F_N A_1 \Psi\,,\quad
A_1=v_1 \cdot \partial_{x_i}\,.
$$
Since $\eta^\prime(y)=\delta(y) $, where
 $\delta(y) $ denotes the Dirac
delta-function, we obtain
\[
\begin{split}
&
A \,\Psi = v_1 \cdot \partial_{x_1}
\prod\limits_{i=2}^N \eta[d^2- |x_1-x_i|^2]=
\\
&
= 2 \, \sum\limits_{i=2}^N \!\!
[v_1 \cdot(x_1-x_i)]\,
\delta[|x_1-x_i|^2 -d^2]\,
{
\prod\limits_{j=2}^{N}
}^{(j \neq i)} \eta[d^2- |x_1-x_j|^2]\,.
 \end{split}
\]
Then we perform the integration in~\eqref{6.9}
and use again symmetry of $F_N$ and $\Psi$.
The result reads
\[
\begin{split}
&I_1= v \cdot \partial_{x_1} F_1^{(N)}(x_1,v_1) -
\\
&- 2(N-1) \!\int\limits_{ \mathbb{R}^3\times \mathbb{R}^3}
\! \! dx_2 dv_2 \,\delta[|x_1-x_2|^2-d^2]
v_1 \cdot (x_1-x_2) \tilde{F}_2^{(N)}
(x_1,v_1;x_2,v_2)
 \end{split}
\]
in the notation of Eq.~\eqref{6.6}. Note that
\[
2 \int\limits_{ \mathbb{R}^3}
 dy \delta[(x-y)^2-d^2]F(y)=
d^2 \int\limits_{S^2} dn F(x-dn).
\]
 Therefore we obtain Eq.~\eqref{6.8} in the following
 form:
 \begin{equation}\label{6.13} 
 \begin{split}
(&
\partial_t + v_1 \cdot \partial_{x_1}) F_1^{(N)}(x_1,v_1)=Q^{(N)}=
\\
&
=
(N-1)\, d^2 \!\!\int\limits_{ \mathbb{R}^3\times S^2}
\!\!dv_2 dn \,
[(v_1-v_2 ) \cdot n]\,\tilde{F}_2^{(N)}
(x_1,v_1; \,x_1 - dn, v_2),
\end{split}
\end{equation}
 where $\tilde{F}_2^{(N)}
(x_1,v_1; x_2,v_2) $ is given in~\eqref{6.6}.
 We can split the integral over $S^2$ into two
 parts in the following way:
  \begin{equation*}
 \begin{split}
 \int\limits_{S^2} dn (u \cdot n) \Psi(n) =
 \int\limits_{S_+^2} dn |u \cdot n| \Psi(n) -
 \int\limits_{S_-^2}  dn |u \cdot n| \Psi(n)\,,
 \\
 S_+^2=\{ n \in  S^2:\, u \cdot n >0  \},
 \quad
 S_-^2=\{ n \in  S^2:\, u \cdot n <0  \},
 \end{split}
\end{equation*}
 where $ u=v_1-v_2,\, \Psi(n) $ is an arbitrary
 integrable function. It is clear from
 Eqs.~\eqref{6.3}, \eqref{6.4}, that
 $ \Psi(n) = \tilde{F}_2^{(N)}
(x_1,v_1; x_1-dn,v_2) $
in the integral over
$  S_+^2 $ can be expressed through $\Psi(n) $ in the integral
 over $S^2_- $.
Then we obtain
\begin{equation}\label{6.14} 
 \begin{split}
 Q^{(N)}= (N-1) \,d^2 \!\!
 \int\limits_{ \mathbb{R}^3\times S^2} \!\!
 dv_2 dn |u \cdot n|\,
 \left[ \tilde{F}_2^{(N)}
 ( x_1,v_1^\prime; \,x_1-dn, \,v_2^\prime ) -\right.
 \\
 \left. -
  \tilde{F}_2^{(N)} (x_1,v_1; \,x_1+dn, \,v_2)
  \right] \,,
  \end{split}
\end{equation}
 \[
 v_1^\prime=v_1- n (u \cdot n), \;
  u=v_1- v_2,
 \;  v_2^\prime = v_2 + n(u \cdot n).
 \]
 Note that
Eqs.~\eqref{6.13}--\eqref{6.14} are formally exact
for hard spheres. To our knowledge, they were
firstly published by Harold Grad not later than in
1957~\cite{Gra}. These equations are very
important as a starting point for mathematically
 rigorous derivation of the Boltzmann equation.

 For our goals it is sufficient to introduce the
 ''chaotic'' initial data
  \begin{equation}\label{6.15}  
 \begin{split}
 F_N(1,2,\dots,N)\big|_{t=0} = c_N
 \left\{\prod\limits_{k=1}^N F_0(k)
 \right\}
 &
 \prod\limits_{1 \leq i < j \leq N}
 \!\! \eta \left[d^2 - |x_i-x_j|^2\right]\,,
 \\
 &\int \limits_{G }d1 \, F_0(1)=1,
  \end{split}
\end{equation}
where $ c_N $ is the normalization constant, and to consider
 the formal limit of Eqs.~\eqref{6.13}--\eqref{6.14} under conditions that
  \begin{equation}\label{6.16} 
N \to \infty, \quad d \to 0, \quad N\,d^2 =
\mathrm{const.}
\end{equation}
It is the so-called Boltzmann-Grad
limit~\cite{Cer1,Gra}. To be more
precise we assume that
  \begin{equation}\label{6.17}
\begin{split}
F_1^{(N)}(x_1,v_1,t) \to F(x,v,t),
\quad
F_1^{(N)}(x_1,v_1; x_2,v_2; t) \to
\prod \limits_{i=1}^2 F(x_i,v_i,t)
\end{split}
\end{equation}
under conditions~\eqref{6.16}.

Then we formally obtain from~\eqref{6.13},
\eqref{6.14} the Boltzmann equation for hard
spheres. It simplified notations
$ x=x_1,\;v+v_1,\;w=v_2 $ this equation reads
  \begin{equation}\label{6.18}    
(\partial_t + v \cdot \partial_{x}) F(x,v,t)=
(Nd^2)  \tilde{Q}(F,F),
\end{equation}

\begin{equation}\label{6.19}
 \begin{split}
&\tilde{Q}(F,F)=
\frac1{2}
\!\!\int\limits_{ \mathbb{R}^3\times S^2}\!\!
\!
dw dn |u \cdot n| [F(x,v^\prime,t) F(x,w^\prime,t)
-F(x,v,t) F(x,w,t)]\,,
\\
&
u=v-w,\quad n \in S^2; \quad
v^\prime=v -(u \cdot n) n, \quad
w^\prime= w + (u \cdot n),
\end{split}
\end{equation}
where the domain of integration $S_+^2$
in~\eqref{6.14} is extended to the whole unit
sphere $ S^2 $ in obvious way. The limiting
initial conditions formally follows
 from~\eqref{6.15}:

\begin{equation}\label{6.20}  
F(x,v)|_{t=0} = F_0(x,v),
\quad \int \limits_G \!\! dx dv  F_0(x,v)=1.
\end{equation}

We consider in this section the case of the whole
space $G= \mathbb{R}^3 \times \mathbb{R}^3 $ in
order to simplify formal calculations. In fact
the same equations~\eqref{6.18}--\eqref{6.20} can
be derived rigorously from $N-$particle dynamics
in case, when all particles are confined in bounded
domain $\Omega$ with reflecting walls (see~\cite{CIP} and
references therein for details).

In our formal derivation of the Boltzmann equation in this section we followed
the Grad's scheme from \cite{Gra}. It should be pointed out that the first mathematically
rigorous proof of validity of the Boltzmann equation for hard spheres was done
by O. Lanford in 1975 \cite{Lan}. The deep and rigorous presentation of the
validation problem for the Boltzmann equation can be found in the book \cite{CIP}.

\subsection{Boltzmann equation for hard spheres and other potentials}

The classical Boltzmann equation is usually
considered not for the probability density $ F(x,v,t) $,
but for the distribution function $ f(x,v,t) $
\begin{equation}\label{2.1.1} 
f(x,v,t) = N \, F(x,v,t).
\end{equation}

The equation for $ f(x,v,t) $ reads
\begin{equation}\label{2.1.2}
\begin{split}
f_t + v \cdot f_x=& {Q}(f,f)=
\\
&=\frac{d^2}{2}
\int\limits_{ \mathbb{R}^3\times S^2}\!\!
\!
dw dn \,|u \cdot n|\,[f(v^\prime) f(w^\prime)-
f(v) f(w)],
\\
u=&v-w,\quad n \in S^2; \quad
v^\prime=v -(u \cdot n) n, \;
w^\prime= w + (u \cdot n),
\end{split}
\end{equation}
where irrelevant arguments $(x,t)$ of $f(x,v,t)$
are omitted in the so-called Boltzmann collision
integral $ Q(f,f) $. Note that $ Q(f,f) $ is a
quadratic with respect to  $f$ operator acting
only on variable $v \in  \mathbb{R}^3\ $. 
The
advantage of Eq.~\eqref{2.1.2} is that it does not
contain number of particles $N$. The Boltzmann collision integral
 can be presented in different forms.
In particular, the following form of $ {Q}(f,f) $
 is very useful:

\begin{equation}\label{2.1.3}  
Q(f,f)
= \frac{d^2}{4}
\!\!\int\limits_{ \mathbb{R}^3\times S^2 }\!\!
\!
dw d\omega |u |\,[f(v^\prime) f(w^\prime)-
f(v) f(w)]\,,
\end{equation}
\[
u = v-w,\quad \omega \in S^2; \quad
v^\prime=\frac{1}{2}(v+w+|u|\omega), \quad
w^\prime= \frac{1}{2}(v+w-|u|\omega).
\]

In order to prove that this integral coincides
with $ Q(f,f)$ from~\eqref{2.1.2} we consider
a simpler integral
\begin{equation}\label{2.1.4}
I(F)= \int\limits_{\mathbb{R}^3} \!\!
dk\;\delta\left(k \cdot u +\frac{|k|^2}{2}\right)
\,F(k)\,,
\end{equation}
where $ u \in \mathbb{R}^3 $,\, $ F(u) $ is a
continuous function. Then the following identity
can be easily proved:

\newtheorem{Lemma}{Lemma}[section]

\begin{Lemma}\label{lem2.1}

\begin{equation}\label{2.1.5}
I(F) =2 \int\limits_{\mathbb{S}^2} \!\!
dn |u \cdot n|\, F [ - 2 (u\cdot n) n] =
|u|  \int\limits_{\mathbb{S}^2} \!\! d \omega
F(|u|\omega-u).
\end{equation}
\end{Lemma}

\newenvironment{Proof} 
{\par\noindent{\bf Proof.}} 
{\hfill$\scriptstyle\blacksquare$} 

\begin{Proof}

 We evaluate the integral $I(F)$ in
spherical coordinates with polar axis directed
along $ u \in \mathbb{R}^3 $. Thus we denote
$ k=rn$,\; $ n \in \mathbb{S}^2 $ in~\eqref{2.1.4}
and obtain
\begin{equation*}
I(F)= \int\limits_{0}^\infty\!\! dr\, r^2
\int\limits_{\mathbb{S}^2 } \!\! dn \,
\delta\left[rn\cdot u + \frac{r^2}{2}
\right]\, F(rn).
\end{equation*}
Since
\begin{equation}\label{2.1.6}
\delta(\alpha x) = \frac{\delta(x)}{\alpha},
\quad \alpha > 0,\; x \in \mathbb{R}\,,
\end{equation}
we change the order of integration and obtain
\begin{equation*}
\begin{split}
I(F)= &\int\limits_{\mathbb{S}^2} \!\! dn
\int\limits_{0}^{\infty}\!\! dr\, r
\delta \left[ \frac{1}{2} (2 n \cdot u + r) \right]
F(rn)=
\\
=& 4 \int\limits_{\mathbb{S_-}^2 }\!\! dn \,
 |n \cdot u| \;F[- 2 (u \cdot n) n],
 \quad \mathbb{S_-}^2 =\{ n \in \mathbb{S}^2:
 u \cdot n < 0 \}.
 \end{split}
\end{equation*}
The integrand is an even function of $n \in S^2 $
and hence the first equality~\eqref{2.1.5}
follows. The second equality is based on change
of variables $k = \tilde{k} - u$ in the
integral~\eqref{2.1.4}. Then we obtain
\[
I(F) = \int\limits_{\mathbb{R}^3 } \!\! dk
\; \delta \left(\frac{|k|^2 - |u|^2}{2}\right)\; F(k-u)
\]
and evaluate this integral in the same way as
above. 

\noindent
This completes the proof of Lemma~\ref{lem2.1}.
\end{Proof}

\vspace{10pt}

Now we can prove the transformation of $ Q(f,f) $
from~\eqref{2.1.2} to~\eqref{2.1.3}. We
 consider~\eqref{2.1.2} and denote
 \begin{equation}\label{2.1.7}                
 F(k)=f(v+k/2) f(w-k/2) - f(v) f(w),
 \end{equation}
 considering $ v $ and $ w $ as fixed parameters.
Then we obtain from~\eqref{2.1.2}
\[ Q(f,f)= \frac{d^2}{2}
 \int\limits_{\mathbb{R}^3 \times S^2}\!\!
dw \, dn \, |u \cdot n| \,F[-2 (u \cdot n)n].
\]
It remains to use the identity~\eqref{2.1.5} and
get the following result:
\[
Q(f,f)= \frac{d^2}{4}
\int\limits_{\mathbb{R}^3 \times S^2}\!\!
dw \, d \omega\, |u| \, F(|u|\omega - u)
\]
in the notation of Eq.~\eqref{2.1.7}. It is easy
to check that  this formula for $ Q(f,f) $ coincides
with~\eqref{2.1.3}. Hence, the equivalence
of~\eqref{2.1.2} and~\eqref{2.1.3} is proved.
Note also that the same identity~\eqref{2.1.5}
leads to the third useful representation of the
collision integral for hard spheres:
\begin{multline}\label{2.1.8}
Q(f,f) =\frac{d^2}{4} \int\limits_{\mathbb{R}^3\times \mathbb{R}^3} dw dk \;
\delta (k \cdot u + |k|^2/2) \times
\\
\times
[ f(v+k/2) f(w-k/2) - f(v) f(w)].
\end{multline}

The physical meaning of the Boltzmann equation
can be better understood by considering
$ Q(f,f) $ in the form~\eqref{2.1.3}. We denote
\begin{equation}\label{2.1.9}                      
\langle f, \psi \rangle = \int \limits_{\mathbb{R}^3}
\!\!dv\, f(v)\psi(v)\,,
\end{equation}
where $\psi(v) $ is an arbitrary function of
velocity  $ v \in \mathbb{R}^3 $ for which the
integral exists. Then we formally obtain
from~\eqref{2.1.2}
\begin{equation}\label{2.1.10}
\partial_t\langle f, \psi \rangle
+ \partial_x \cdot
\langle f, v\psi \rangle=
\langle \psi, Q(f,f) \rangle.
\end{equation}
The right hand side with $ Q(f,f) $ from~\eqref{2.1.3}
reads
\begin{equation*}
\langle \psi, Q(f,f) \rangle =
  \frac{d^2}{4} \int\limits_{\mathbb{R}^3 \times \mathbb{R}^3
 \times \mathbb{S}^2}\!\!\!\!
 dv dw d \omega\, |u| \psi(v) [f(v^\prime)
 f(w^\prime) - f(v) f(w)]
\end{equation*}
in the notation of~\eqref{2.1.3}. We denote the
center of mass variables (see Section 1.4) by
\begin{equation}\label{2.1.11}                     
U=\frac{v+w}{2}, \quad u =v-w \Leftrightarrow
v=U +\frac{u}{2},\quad w=U -\frac{u}{2}.
\end{equation}
Hence, $ dv dw = dU du $. Therefore we obtain
\begin{equation*}
\begin{split}
\langle \psi, Q(f,f) \rangle =
  \frac{d^2}{4} \int\limits_{\mathbb{R}^3 \times \mathbb{R}^3
 \times \mathbb{S}^2}\!\!\!
 dU du d \omega\, |u| \psi\left(U+\frac{u}{2}\right)
 [F(U,|u|\omega)
 -F(U,u)],\\
 F(U,u)=f(v)f(w).
 \end{split}
\end{equation*}
If we denote $u=r \omega_0, \; \omega_0 \in S^2$,
and write down the integral over $ du $ as
\[
\int\limits_{\mathbb{R}^3} du \varphi(u)
 = \int\limits_0^\infty
dr \, r^2
\int\limits_{\mathbb{S}^2} d\omega_0
\varphi(r\omega_0),
\]
then the internal integral over $ d\omega_0 d\omega $
reads
\[
I=\int\limits_{{S}^2\times S^2}
d\omega_0 d\omega
\psi\left(U+\frac{r}{2}\omega_0\right)
[F(U,r \omega) - F(U, r \omega_0)],
\]
where $r= |u|$. Obviously we can exchange variables
$\omega$ and $\omega_0 $ in the first term and
obtain
\[
I=\int\limits_{{S}^2\times S^2}
d\omega_0 d\omega
F(U,r \omega_0)\,\left[ \psi(U+\frac{r}{2} \omega)
 - \psi(U+\frac{r}{2}\omega_0)
\right]\,.
\]
Coming back to initial variables, we get
\begin{equation}\label{2.1.12}
\langle \psi, Q(f,f) \rangle =
\int\limits_{\mathbb{R}^3 \times \mathbb{R}^3 }
\!\! dv dw f(v) f(w) |u| \,\sigma_{tot}\,
\overline{[\psi(v^\prime)-\psi(v)]},
\end{equation}
where
\begin{equation}\label{2.1.13}                      
\begin{split}
 &
 \sigma_{tot}=\int\limits_{\mathbb{S}^2}
 d\omega \sigma_{diff} = \pi d^2,
 \\
 &
\overline{\psi(v^\prime) -\psi(v) }= \frac{1}
 {\sigma_{tot}} \int\limits_{\mathbb{S}^2}
  d\omega \sigma_{diff}
  \left[ \psi\left( U+
  \frac{|u|}{2} \omega \right) -
  \psi\left( U+
  \frac{u}{2} \right) \right]\,,
\end{split}
\end{equation}
$ \sigma_{diff}=d^2/4 $ denotes the so-called differential
scattering cross-section for hard shperes with diameter  $ d $.
 The bar in~\eqref{2.1.12}, \eqref{2.1.13} means
actually an averaging over random
impact parameters. The
physical meaning of Eq.~\eqref{2.1.10} becomes clear
if we write it as
\begin{equation}\label{2.1.14}                   
\partial  _t\langle f, \psi \rangle  +
 \partial_x \cdot \langle f, v\psi\rangle=
\int\limits_{\mathbb{R}^3\times \mathbb{R}^3}
dv dw f(v) f(w) |u|\sigma_{tot}
[\overline{\psi(v^\prime) -\psi(v)}]
\end{equation}
in the notation of Eq.~\eqref{2.1.13}. Indeed the
average total  number of collisions per unit time
is given by the integral
\begin{equation}\label{2.1.15}  
\nu_{tot}(f,f)=\int\limits_{\mathbb{R}^3\times \mathbb{R}^3}
dv dw f(v) f(w) |u|\sigma_{tot}\,.
\end{equation}
On the other hand, the average change  of $ \psi $
in the collision of particles with velocities
 $ v $ and $ w $ is equal to the average difference
  $ [\overline{\psi(v^\prime) -\psi(v)}] $ given 
  in Eq.~\eqref{2.1.13}.
 Hence, the right hand side of Eq.~\eqref{2.1.14}
 defines correctly (at the intuitive level) the
 rate of change of $ \langle f, \psi \rangle $
 due to collisions.

These considerations allows us to generalize
Eqs.~\eqref{2.1.12}--\eqref{2.1.15} to the case
of general (repulsive) potential $ \Phi(r)$ with
 finite radius of action $ R_{\max} =d $. In that
 case we have the same total cross-section
 $ \sigma_{tot}=\pi d^2 $, as for hard spheres with
 diameter $d $. However, the differential
 cross-section
 of scattering $(v,w) \to (v^\prime, w^\prime) $,
 such that
 \[
 v-w=u, \quad v^\prime - w^\prime = u^\prime=
 |u| \omega,
 \]
is given for the general potential $ \Phi(r) $
by the function
\begin{equation}\label{2.1.16}                         
\sigma_{diff }= \sigma(|u|, \omega \cdot u / |u|),
\quad \sigma_{tot} = 2 \pi \int \limits_{-1}^{1}
d \mu \sigma(|u|, \mu )
\end{equation}
discussed in detail in textbook in classical mechanics (see \cite{LL1}).
The connection of $  \sigma_{diff}(|u|, \mu)  $ with the  potential
$ \Phi(x) $ is briefly discussed below in Section 2.8.
If we fix the intermolecular potential $ \Phi (r) $
and the corresponding differential scattering
cross-section $ \sigma(|u|, \mu) $   then we obtain the
same equation~\eqref{2.1.12}, where
\begin{equation}\label{2.1.17}
\begin{split}
\overline{\psi( v^\prime) - \psi(v)} =
\frac{1}{\sigma_{tot}} \int\limits_{{S}^2}d\omega \,
\sigma(|u|, \omega \cdot u/ |u|)
\,[ \psi( v^\prime) - \psi(v)  ]\,,
\\
v^\prime = \frac{1}{2} (v+w+ |u| \omega ),
\quad u = v - w \,.
\end{split}
\end{equation}
Finally we note that the total cross-section
$\sigma_{tot} $ disappears after substitution
of~\eqref{2.1.13} into \eqref{2.1.12}.
In addition, the differential  cross-section
$ \sigma(|u|, \mu ) $ is always multiplied by
$ |u| $ in the integrand in~\eqref{2.1.12}.
Therefore it is more convenient to introduce  a
new function
 \begin{equation}\label{2.1.18}                  
g(|u|, \mu ) = |u| \sigma (|u|, \mu ),
\quad \mu \in [-1,1].
\end{equation}
Then the general Eq.~\eqref{2.1.12} reads

 \begin{equation}\label{2.1.19}
\begin{split}
\langle \psi,& Q(f,f) \rangle =
 \\
&=  \!\!\!\int\limits_{\mathbb{R}^3 \times  \mathbb{R}^3
 \times {S}^2} \!\!\!\! \! dv dwd\omega f(v) f(w)
 g(|u|, \omega \cdot u / |u|) \,
 [\psi(v^\prime) - \psi(v)],
 \\
&
u=v-w,\quad v^\prime = \frac{1}{2} [v+w+
|u| \omega], \quad \omega \in S^2\,.
\end{split}
\end{equation}
Note that $\psi(v) $ is an arbitrary test function.
Therefore this expression is sometimes called
''a weak form of the Boltzmann collision integral''.
The corresponding strong form of $ Q(f,f) $ is

 \begin{equation}\label{2.1.20}                           
\begin{split} 
&
Q(f,f)=\!\! \int\limits_{\mathbb{R}^3 \times S^2}\!\!
\!\!dw \, d \omega\
 g(|u|, \omega \cdot u / |u|) [f(v^\prime) f(w^\prime)-
 f(v)f(w)]\,,
 \\
 &
 \omega \in S^2,\;u=v-w,\; v^\prime =
 \frac{1}{2} (v+w + |u| \omega), \;
 w^\prime =
 \frac{1}{2} (v+w - |u| \omega),
\end{split}
\end{equation}
where $ g ( |u|, \mu)$ is given in~\eqref{2.1.18}.
We consider below the Boltzmann equation
\begin{equation}\label{2.1.21}                             
f_t + v \cdot f_x = Q(f,f)\, ,
\end{equation}
in the notation of Eqs.~\eqref{2.1.18}, \eqref{2.1.20}.
 We shall usually consider  $ g ( |u|, \mu)$
as a given function.



\subsection{Basic properties of the Boltzmann equation}

In applications to rarefied gas dynamics we are mainly
interested not in the distribution function $f(x,v,t)$
itself, but in the (macroscopic) characteristics of the
gas averaged over the velocity space. In accordance with
physical meaning of $ f(x,v,t) $, the density of the gas
or equivalently the number of particles per unit volume
is defined by equality
\begin{equation}\label{2.2.1}
\rho(x,t)=\langle  f,1 \rangle=
\int\limits_{\mathbb{R}^3}
dv f(x,v,t),\quad x \in\mathbb{R}^3,\;t \geq 0.
\end{equation}

Another important macroscopic characteristics of the
gas are the bulk or mean velocity $ u(x,t) $ (not
to be confused with the notation $ u $ for relative
velocity in the collision integral \eqref{2.1.20})
and the absolute temperature $ T(x,t) $. These
functions are defined by equalities
\begin{equation}\label{2.2.2}                              
u(x,t)= \frac{1}{\rho} \langle f, v \rangle,\quad
T(x,t)=\frac{m}{3 \rho} \langle f, |v - u |^2 \rangle,
\end{equation}
in the notation of~\eqref{2.1.3}.  Here $ m $ stands
for molecular mass, whereas $ T $ is expressed in energy
 units. Usually we assume in this book that $ m=1 $
 unless the mixture of different gases is considered.

 For given values of
$ \rho(x,t) $,   $ u(x,t) $ and  $ T(x,t) $ the following
distribution function will be called a local Maxwell
distribution (or Maxwellian):

\begin{equation}\label{2.2.3}
f_M(x,v,t)=  \rho ( 2 \pi T)^{-3/2}
\exp\left[- \frac{|v-u|^2}{2 T}\right],
\end{equation}
where it is assumed that $ m=1 $ in~\eqref{2.2.2}.
The same function is called ''absolute Maxwellian''
if the parameters $ \rho, u, T $ are independent of $ x $
and $ t $.

Coming back to the Boltzmann equation~\eqref{2.1.21}
we can easily understand the importance of the
Maxwellian distribution~\eqref{2.2.3}. Indeed it
follows from~\eqref{2.1.20} that

\begin{equation}\label{2.2.4}
v^\prime + w^\prime= v+w, \quad
|v^\prime|^2 + |w^\prime|^2= |v|^2+|w|^2\,,
\end{equation}
i.e. the conservation laws for momentum and energy
in each pair collision. Hence, any function of the
form
\begin{equation}\label{2.2.5}                            
f(v)= \exp(\alpha+ \beta \cdot v - \gamma |v|^2),
\quad \gamma >0,
\end{equation}
with constant parameters $ (\alpha,\beta,\gamma) $
satisfies equations

\begin{equation}\label{2.2.6}
f( v^\prime)f(w^\prime)=f(v)f(w),\quad
v \in \mathbb{R}^3,\quad w \in \mathbb{R}^3,
\quad \omega \in S^2,
\end{equation}
in the notation of Eqs.~\eqref{2.1.20}.  Hence,
\begin{equation}\label{2.2.7}                    
Q(f_M,f_M)= 0
\end{equation}
for any local Maxwellian~\eqref{2.2.3}.

Another important property of the Boltzmann equation
is connected with conservation laws for mass, momentum
and energy. We consider the identity~\eqref{2.1.18} for
a given test function $\psi(v)$ and transform the
integral by exchanging variables $v$ and $w$. Then
we easily obtain
\begin{equation}\label{2.2.8}
\begin{split}
\langle \psi, Q(f,f)\rangle = \frac{1}{2}
\!\! \int \limits_{ \mathbb{R}^3 \times \mathbb{R}^3
\times S^2 } \!\!\! \!dv dw d\omega f(v) f(w)
g(|u|,\omega \cdot u / |u|)[\psi(v^\prime) +
\\
+
\psi(w^\prime) -\psi(v) -\psi(w)]
\end{split}
\end{equation}
in the notation of~\eqref{2.1.20}. Hence,

\begin{equation}\label{2.2.9}                              
\langle \psi, Q(f,f)\rangle = 0 \quad
\text{if} \quad \psi(v)=a + b\cdot v+ c|v|^2
\end{equation}
with any constant parameters $a,b,c$. This
identity leads to conservation laws for mass,
momentum and energy. Indeed we consider Eq.~\eqref{2.1.10}
with $\psi=1$, $\psi =v$ and $\psi =|v|^2$
respectively. Then we obtain
\begin{equation}\label{2.2.10}
\begin{split}
&
\partial_t \rho + \mathrm{div} \rho u =0,
\\
&\partial_t \rho u_\alpha + \partial_{x_\beta}
\langle f, v_\alpha v_\beta \rangle =0,\quad \alpha,\beta=1,2,3;
\\
&
\partial_t \langle f, |v|^2 \rangle+
\mathrm{div}\langle f, |v|^2  v \rangle=0.
\end{split}
\end{equation}
These equations are very basic for the Boltzmann equation and
transition to hydrodynamics.

Finally we shall prove the famous Boltzmann's
$H-$~theorem. The theorem is based on the following
inequality:
\begin{equation}\label{2.2.11}                                 
\langle \log f, Q(f,f) \rangle  \leq 0\,.
\end{equation}

To prove this we need one more identity for
$ \langle \psi, Q(f,f)\rangle   $, namely,

\begin{equation}\label{2.2.12}
\begin{split}
\langle \psi, Q(f,f)\rangle = -  \frac{1}{4}
\!\! \int \limits_{ \mathbb{R}^3 \times \mathbb{R}^3
\times S^2 } \!\!\! dv dw d\omega
g(|u|,\omega \cdot u / |u|)[f(v^\prime) f(w^\prime)
-
 \\
-f(v) f(w)][\psi(v^\prime) + \psi(w^\prime)-
\psi(v)- \psi(w)]\,.
\end{split}
\end{equation}
It follows from Eq.~\eqref{2.2.8} and another general
equality
\begin{equation}\label{2.2.13}                                
\!\! \int \limits_{ \mathbb{R}^3 \times \mathbb{R}^3
\times S^2 } \!\!\! dv dw d\omega
[\Psi(v,w; v^\prime,w^\prime) -
\Psi(v^\prime,w^\prime; v,w )]=0\,,
\end{equation}
valid for any function $ \Psi(v_1, v_2; v_3, v_4) $
such that the integral is convergent. For the proof
it is sufficient to pass to variables
$ U, u $~\eqref{2.1.11}  in the integrand and to repeat
the considerations used for the proof of identity~\eqref{2.2.8}.
For brevity we omit these
straightforward calculations.

To complete the proof of~\eqref{2.2.12} we consider
Eq.~\eqref{2.2.8} and  denote
\[
\begin{split}
\Psi(v,w;\,v',w')=\frac1 2 \,f(v) f(w) \,g(|u|,u^\prime \cdot u / |u|^2)
[\psi(v^\prime) + \psi(w^\prime)-
\psi(v)- \psi(w)]\,,
\\
u^\prime =
 v^\prime- w^\prime=
|u| \omega, \quad u=v-w.
\end{split}
\]
Then we apply~\eqref{2.2.13} and obtain
\[
\langle \psi, Q(f,f)\rangle =   \frac{1}{2}
\!\! \int \limits_{ \mathbb{R}^3 \times \mathbb{R}^3
\times S^2 } \!\!\! dv dw d\omega
[\Psi(v,w; v^\prime,w^\prime) -
\Psi(v^\prime,w^\prime; v,w )]\,,
\]
i.e. the identity~\eqref{2.2.12}. Now we can prove
the inequality~\eqref{2.2.11} by substitution
 of $ \psi = \log f(v) $ into~\eqref{2.2.12}. We obtain
 \begin{equation}\label{2.2.14}                               
\begin{split}
\langle \log f, Q(f,f)\rangle = -  \frac{1}{4}
\!\! \int \limits_{ \mathbb{R}^3 \times \mathbb{R}^3
\times S^2 } \!\!\! dv dw d\omega
g(|u|,\omega \cdot u / |u|)[f(v^\prime) f(w^\prime)
-
 \\
-f(v) f(w)]  \log\frac{f(v^\prime)f(w^\prime)}{f(v)f(w)}
\,\leq 0\,.
\end{split}
\end{equation}
This completes the proof of inequality~\eqref{2.2.11}.

The main application of~\eqref{2.2.11} is the proof of
Boltzmann's $ H-$~theorem. We introduce the
Boltzmann's $ H-$~functional
 \begin{equation}\label{2.2.15}
H(f)(x,t) = \langle f, \log f\rangle=
\int\limits_{\mathbb{R}^3}\!\! dv f(x,v,t) \log f(x,v,t)\,,
\end{equation}
where $f(x,v,t) $ is a solution of Eq.~\eqref{2.1.21}.

Note that
\[
(\partial_t + v \cdot \partial_x ) f \log f=
(1+ \log f) (f_t + v \cdot f_x)=
(1+ \log f) Q(f,f)\,.
\]
Hence, we obtain by integration in $v$
 \begin{equation}\label{2.2.16}
\partial_t \langle f, \log f \rangle +
\mathrm{div} \langle f, v \log f\rangle =
\langle \log f , Q(f,f)\rangle \,\leq 0
,.
\end{equation}
This inequality is known as ''the Boltzmann's
 $H-$~theorem''. Its importance can be easily understood in
 the spatially homogeneous case, considered in the
 next section.

 \subsection{Spatially homogeneous problem}

 The Boltzmann equation~\eqref{2.1.21} admits
  a class of spatially homogeneous solutions $f(v,t)$. We
 usually consider the initial value problem
  \begin{equation}\label{2.3.1}                                
  f_t=Q(f,f),\quad f|_{t=0} = f_0(v)\,,
\end{equation}
in the notation of Eq. \eqref{2.1.20}.

The conservation laws~\eqref{2.2.10} show that
  \begin{equation}\label{2.3.2}
\begin{split}
\rho =\langle f,1\rangle =\mathrm{const.},\;
u= \frac{1}{\rho} \langle f,v\rangle=\mathrm{const.},
\\
T=\frac{1}{3 \rho} \langle f, |v-u|^2\rangle =
\frac{1}{3 \rho} [\langle f, |v|^2 \rangle - \rho |u|^2]
=\mathrm{const.}
\end{split}
\end{equation}
We also note that the operator $Q(f,f)$ is invariant
under shifting $ v \to v +v_0$,
$v_0 \in \mathbb{R}^3 $, in $v-$space. Therefore if
$f(v,t)$ is a solution of the equation from~\eqref{2.3.1},
then $f(v+v_0, t)$ is also a solution for any
$v_0 \in \mathbb{R}^3 $. Hence, we can always reduce
the problem~\eqref{2.3.1} to the case
\begin{equation}\label{2.3.3}                                 
u = \frac{1}{\rho} \langle  f, v\rangle =0,
\quad t \geq 0.
\end{equation}
Moreover if $f(v,t) $ is a solution of the spatially
homogeneous Boltzmann equation, then so is the
function $ \tilde{f}(v,t) = \alpha f(v, \alpha t) $
with any $  \alpha >0 $. This transformation allows
to reduce the general problem~\eqref{2.3.1} to the
case
\begin{equation}\label{2.3.4}
\rho = \langle  f,1 \rangle =1.
\end{equation}
The the corresponding Maxwell distribution~\eqref{2.2.3}
reads
  \begin{equation}\label{2.3.5}                        
f_M(v) = (2 \pi T)^{3/2} \exp\left( -
\frac{|v|^2}{2T} \right),
\end{equation}
where
\begin{equation}\label{2.3.6}                                    
T= \frac{1}{3} \langle  f_0, |v|^2 \rangle.
\end{equation}
The $H-$theorem~\eqref{2.2.16} shows that
 the functional
\begin{equation}\label{2.3.7}
H(f) = \langle f, \log f \rangle =
\int\limits_{\mathbb{R}^3} dv f(v,t) \log f(v,t)
 \end{equation}
cannot increase with time on the solution of~\eqref{2.3.1}
 because
\begin{equation}\label{2.3.8}
 \partial_t H(f)(t) = \langle \log f, Q(f,f)\rangle
 \leq 0.
  \end{equation}
 If we consider the explicit formula~\eqref{2.2.14},
 then it becomes clear that $ \langle f, Q(f,f)\rangle
 =0 $ if and only if
 $$f(v^\prime) f(w^\prime)=f(v) f(w)  $$
  for almost all values $(v,w, \omega)  \in \mathbb{R}^3
  \times \mathbb{R}^3\times S^2$ provided
  the kernel  $ g(|u|, \mu ) $
  in~\eqref{2.1.20} is positive almost
  everywhere. This functional equation was studied
  in various classes of functions by many authors
  beginning with L. Boltzmann (see~\cite{CIP} and
  references therein). They proved the uniqueness of
  its well-known solution~\eqref{2.2.5}. On the
  other hand, the only function~\eqref{2.2.5}, which
  satisfies the above discussed conservation laws is
  the Maxwellian $f_M$~\eqref{2.3.5}. Hence, we can
  conclude at the formal level, that $ H(f) $
  decreases monotonically in time unless $f=f_M$.
  This conclusion can be confirmed by general inequality
  \begin{equation}\label{2.3.9}
\langle f_M,\log f_M \rangle \leq
 \langle  f, \log f\rangle
\end{equation}
in the notation of Eqs.~\eqref{2.2.1}--\eqref{2.2.3}.
Its proof is very simple. Note that
$\langle f-f_M,\log f_M \rangle=0 $. Therefore it
is sufficient to prove that
$\langle  f, (\log f-\log f_M) \rangle  \geq 0 $.
This follows from elementary inequality
\[
G(z,y) = z ( \log z - \log y) + y -z =
z G_1\left( \frac{z}{y}\right) \geq 0,
\quad z>0,\,y>0,
\]
where $ G_1(t) = \log t +t^{-1} -1 \geq 0 $.
We set $z =f(v), \, y= f_M(v) $ and  integrate the
inequality $G(f,f_M) \geq 0 $ over $ v \in \mathbb{R}^3 $.
This completes the proof of~\eqref{2.3.9}. The
inequality~\eqref{2.3.9} shows that the
Maxwellian~\eqref{2.2.1}--\eqref{2.2.3} is the minimizer
of the $H-$functional
$H(f) = \langle f, \log f \rangle\rangle $
in the class of distribution functions with fixed lower
moments $(\rho , u, T)$.

Coming back to the initial value problem~\eqref{2.3.1}
and assuming conditions~\eqref{2.3.3}, \eqref{2.3.4},
we know that there is a unique positive stationary
solution $f_M$, given in~\eqref{2.3.5}, \eqref{2.3.6},
such that all conservation laws are satisfied. This
stationary solution $ f_M $ minimizes   $H-$functional
and therefore we expect that the solution $f(v,t)$
converges (in some precise sense) to $f_M$ for large
values of time.

This is a qualitative behavior of solutions of the
problem~\eqref{2.3.1} that we expect on the basis of
above formal considerations. The corresponding
physical process is called ''the relaxation to
equilibrium''.

Rigorous mathematical theory of the problem~\eqref{2.3.1}
is not simple. First steps in its development were
made by T. Carleman in 1930s \cite{Car} for the model of
hard spheres. Then a more general and detailed theory
was presented by L. Arkeryd~\cite{Ar1} in
early 1970s (see also~\cite{Ar2}). To understand these and more recent results
 in this area we need to introduce a sort of
classification of collisional kernels $g(|u|, \mu)$ in
the Boltzmann integral~\eqref{2.1.20}. This is done in
the next section.

\subsection{Collisional kernels}

We remind to the reader that the kernel $ g(|u|, \mu) $
is equal to $ |u| $ multiplied by the differential
 cross-section $\sigma(|u|,\mu)$ expressed as  a
 function of $\mu=\cos \theta $. The scattering angle
  $ \theta \in [0, \pi ] $ is given in the form
  (see \cite{LL1})

  \begin{equation}\label{2.4.1}
\theta (b, |u|) = \pi - 2 b |u|
\int\limits_{r_{\min}}^\infty\!\!
\frac{r}{r^2}
\left[ |u|^2 \left( 1- \frac{b^2}{|u|^2}\right)
-\frac{2 \Phi(r)}{m}\right]^{-1/2},
\end{equation}
where $b$ is the impact parameter, $\Phi(r)$ is
intermolecular potential, $m$ is the reduced mass
of colliding particles ($m=1/2 $ for particles with
unit mass). To find $ \sigma(|u|, \mu) $ we need to
to construct the inverse function $ b=(|u|, \theta) $.
Then we express this function as
 $b =\tilde{ b}(|u|, \cos \theta) $ and finally
 obtain (see \cite{LL1})
 \begin{equation}\label{2.4.2}                           
\sigma(|u|, \mu)=\left| \frac{1}{2}
\partial_\mu \tilde{ b}^2(|u|, \mu) \right|.
\end{equation}
Generally speaking, this is a rather complicated
calculation. Fortunately it leads to a simple
explicit formula $\sigma = d^2/4 $ in the important
case, when particles are hard spheres with diameter
$ d$. It also can be that for
power-like potentials $ \Phi(r) = \alpha r ^{-n} $,
 $\alpha > 0 $, we obtain
 \[
 \sigma(|u|, \mu)= \left( \frac{\alpha}
 {m |u|^2} \right)^{2/n} \tilde{A}_n(\mu),
 \quad \mu = \cos \theta,\; n \geq 1,
 \]
where a function of $ theta $ is expressed as the
function   $  \tilde{A}_n  $  of $\mu $. Hence, in the case of
power-like potentials $ \Phi(r) = \alpha r ^{-n} $
the collisional kernel in~\eqref{2.1.20} reads
\begin{equation}\label{2.4.3}
g(|u|, \mu)= |u|^{\gamma_n} g_n(\mu),
\quad {\gamma_n}=1 - 4/n\,.
\end{equation}

We can use the same formula for hard spheres assuming
that $n = \infty, \; g_\infty =d^2/4 $. There is, however,
an important difference between hard spheres and
power-like potentials. We consider again the collision
integral~\eqref{2.1.20} and split it formally into
two parts:
\begin{equation}\label{2.4.4}
Q(f,f) = Q^{gain}(f,f) - Q^{loss}(f,f)\,,
\end{equation}
where
\begin{equation}\label{2.4.5}
\begin{split}
Q^{loss}(f,f)=f(v)\nu(v), \quad
\nu(v)=\int\limits_{\mathbb{R}^3} \!\!
dw f(w) g_{tot}(|v-w|)\,,
\\
g_{tot}(|u|) = | u| \sigma_{tot}(|u|) =
2 \pi |u| \int \limits_{-1}^{1} \!\!d\mu \sigma(|u|,\mu).
\end{split}
\end{equation}
It was already discussed in Section 1.5 that
$\sigma_{tot} = \pi R^2_{\max} $, where $R_{\max}$
denotes the radius of action of the potential.
In case of hard spheres of diameter $d$ or any
potential with $ R_{\max} = d $ we obtain the universal
formula for the collision frequency:
  \begin{equation}\label{2.4.6}                              
\nu(v) = \pi d^2   \int\limits_{\mathbb{R}^3}
dw f(w) |v -w|\,.
      \end{equation}

However, if we consider the power-like potential with
any $ n > 0 $, then $ R_{\max} = \infty $ and
therefore the integral $  \nu(v) $ diverges.
Hence, the splitting \eqref{2.4.4} is impossible,
though the ''whole'' collision integral~\eqref{2.1.20}
can be convergent. The matter is that the kernel
$ g(|u|, \mu) $ has a non-integrable singularity at
$ \mu=1 $, i.e. $\theta=0 $, for long range potentials with
$ R_{\max} = \infty $. At the same time $v^\prime = v$ and
$w^\prime =w $ if $ \mu = 1 $. Therefore the second
factor in the integrand is equal to zero at that point.
It is easy to see that the integral~\eqref{2.1.20} is
convergent for a large class of functions $ f(v) $
provided
\begin{equation}\label{2.4.7} 
 \int \limits_{-1}^{1} \!\!d\mu g_n(\mu) (1 - \mu)
 < \infty
      \end{equation}
in case of power-like potentials.
It can be shown that this condition is satisfied for
all $ n > 1 $. The  Coulomb case $  n=1  $ is always
considered separately.

There are also many publications in last two decades related to Boltzmann
equation with
long range potentials, but we do not consider related problems below (see e.g.
\cite{Ale, Morg} and references therein). Our main goal is to discuss a general class
of Boltzmann-type equations, which are used in by physicists. This class of
equations is introduced in the next section.


\section{Boltzmann-type kinetic equations and their \\ discrete models}


\subsection{Generalization of the Boltzmann equation}
\numberwithin{equation} {section}

We introduce in this section a general class of kinetic equations that includes the spatially homogeneous 
Boltzmann equation  \eqref{2.3.1} as a particular case. To this goal we
choose a function $ F(x_1,x_2 ; x_3,x_4)  $ of four
real (or complex) variables   and assume that
\begin{equation}\label{eq-3.1}
		F(x_1,x_2 ; x_3,x_4) = F(x_2,x_1 ; x_3,x_4) = F(x_1,x_2 ; x_4,x_3) = -F(x_3,x_4 ; x_1,x_2).
	\end{equation}

We also introduce a non-negative function $ R(u_1,u_2) $	of two
vectors  $ u_1, u_2 \in \R^3 $.
It is usually assumed that $ R(u_1,u_2) $ is invariant under rotations
of  $ \R^3 $. Then it is known that
	\begin{equation}\label{eq-3.2}
		R(u_1, u_2) = R_1(|u_1|^2, |u_2|^2, u_1 \cdot u_2),
	\end{equation}
i.e. such function can be reduced to a function  of three
scalar variables. This property holds for $ d $-dimensional
vectors with any $ d \ge 2 $.	The function $ R(u_1, u_2) $ will play below a role of the kernel of certain
integral operator.

Then we define the general Boltzmann-type kinetic equation
for a function $ f(v,t) $, where $ v \in  \R^3 $,
$ t \in \R_+ $,  by equality 
\begin{equation}\label{eq-3.3}
		f_t(v,t) = K[f](v),
	\end{equation}
where the general kinetic operator $ K $	acts on
$ v $-variable only.  It is defined by formula
		\begin{multline}\label{eq-3.4}
		K[f](v) = \! \!\int \limits_{\R^3 \times \R^3 \times \R^3} \!\!\! \! \! \!
		 dv_2 dv_3 dv_4
		  \,\delta[v + v_2 - v_3 - v_4]
		   \,\delta\left[|v|^2 + |v_2|^2 - |v_3|^2 - |v_4|^2 \right]\times
		  \\
		\times  {R}(v - v_2, v_3 - v_4) F[f(v), f(v_2); f(v_3), f(v_4)].
	\end{multline}
Our nearest goal is to simplify this integral for
arbitrary $ F(x_1,x_2 ; x_3,x_4)  $ and, in particular, to show that
$ K[f] $ coincides with the Boltzmann collision integral if
\begin{equation}\label{eq-3.5}
		F(x_1,x_2,x_3,x_4) = x_3 x_4 - x_1 x_2.
	\end{equation}
We note that for any $ \alpha \neq 0 $,
\begin{equation*}
\delta(\alpha x) = | \alpha |^{-1}\delta(x),
\quad
|v|^2 + |v_2|^2 - |v_3|^2 - |v+v_2-v_3|^2 = -2 (v_3-v_2)
\cdot(v_3 - v).
\end{equation*}
Hence,
\begin{gather}
		K[f](v) = \frac12 \int \limits_{\R^3 \times \R^3} dv_2\, dv_3\, \delta \left[(v_3 - v_2) \cdot (v_3 - v) \right] \times
		\notag\\
		\qquad \qquad \times \, R(v - v_2, 2v_3 - v - v_2) \;\;
		 F[f(v), f(v_2);\,f(v_3), f(v+ v_2 -v_3)].
		  \label{eq-3.6}
	\end{gather}
Changing variables by formulas
\begin{equation*}
v_2=w, \quad
v_3= v + \frac{k}{2},
\end{equation*}
we obtain
\begin{multline}
		K[f](v) = \frac18 \int \limits_{\R^3 \times \R^3} dw\, dk\, \delta (k \cdot u + |k|^2 / 2) R( u, u + k) \times\\
		\times F[f(v),f(w);  f(v + k / 2),f( w - k / 2)],\quad u = v - w. \label{eq-3.7}
	\end{multline}
Then we use formulas \eqref{2.1.4}, \eqref{2.1.5}  and obtain after simple calculations	
\begin{equation}\label{eq-3.8}
		K[f](v) = \frac18 \int \limits_{\R^3 \times S^2} dw\, d\omega\, |u| R(u, |u| \omega) F[f(v), f(w); f(v'), f(w')],
	\end{equation}	
where	
\begin{equation}\label{eq-3.9}
\omega \in S^2,
\quad
u = v - w,\quad v' = (v + w +u') / 2,\quad 
u'= |u| \omega, \quad  w' = (v + w -u') / 2. 	
\end{equation}		
This formula for	$ K[f](v)  $	is valid for any
 kernel $ R(u_1, u_2)$.  In case of isotropic (invariant under rotations) kernel we can use~\eqref{eq-3.2} and denote
\begin{equation}\label{eq-3.10}	
	R(u, |u| \omega) = 8 |u|^{-1} g(|u|, \hat{u}\cdot \omega),
	\quad
\hat{u} = \frac{u}{|u|}.	
\end{equation}		
Then we obtain
	\begin{gather}
		K[f](v) = \int \limits_{\R^3 \times S^2} dw\, d\omega\, g(|u|, \hat{u} \cdot \omega) F[f(v), f(w); f(v'), f(w')],\label{eq-3.11}
	\end{gather}
where all notations are the same, as in \eqref{2.1.20}.
If, in addition, we assume that $ F $	
 is given by~\eqref{eq-3.5}, then $ K[f](v)  $ coincides
with the Boltzmann collision integral	\eqref{2.1.20}.
Note that the simplest case of constant kernel
$  R(u_1, u_2) = const. $ in~\eqref{eq-3.4} corresponds, under condition \eqref{eq-3.5},
 to the case of hard
spheres for the Boltzmann equation .
 Thus the following
statement is proved.
\vspace{10pt}

\newtheorem{Prop}{Proposition}
\begin{Prop}\label{Prop3.1}	
The equation~\eqref{eq-3.3}, \eqref{eq-3.4}	with 
isotropic kernel~\eqref{eq-3.2} can be reduced by formal
transformations to the Boltzmann-type equation \eqref{eq-3.3}, \eqref{eq-3.11}. The connection between kernels of
corresponding integral operators is given by equality~ \eqref{eq-3.10}. If $ F(x_1,x_2 ; x_3,x_4)  $ in
operator $ K $~\eqref{eq-3.4} is given by formula~ \eqref{eq-3.5}, then the equation~\eqref{eq-3.3}, \eqref{eq-3.4} coincides  with the spatially homogeneous
Boltzmann equation \eqref{2.3.1}, \eqref{2.1.20}.	
		
\end{Prop}

Some authors (see e.g. \cite{DymK}) consider a $ d $-dimensional version of integral \eqref{eq-3.4}, where
$ \R^3 $ is replaced by $ \R^d $, $ d \geq 2 $. Then all
above transformations can be repeated for the integral
$ K[f](v)  $, $ v \in \R^d $, with minimal changes \cite{BK}. The $ d $-dimensional analogue of 
equality \eqref{eq-3.8} reads
\begin{equation}
\label{eq-3.12}
		K[f](v) = 2^{-d}  \! \!\int \limits_{\R^3 \times S^{d-1}} \! \! dw\, d\omega\, |u|^{d-2} R(u, |u| \,\omega)  \;F[f(v), f(w); f(v'), f(w')],
	\end{equation}
in the notation of \eqref{eq-3.9}, where 
$ v \in \R^d $, $ w \in \R^d $, $ \omega \in S^{d-1} $.
If we assume that  the kernel $ R(u_1, u_2) $ is invariant under rotations of $ \R^d  $, then we can use
the same identity~\eqref{eq-3.2} and denote
\begin{equation}\label{eq-3.13}
		R(u, |u| \omega) = 2^d |u|^{2-d} g(|u|, \hat{u} \cdot \omega),\quad
		\hat{u} = \frac{u}{|u|}.
	\end{equation}
The substitution of the formula \eqref{eq-3.13} into \eqref{eq-3.12}
leads to the $ d-$-dimensional "collision integral"        \,\eqref{eq-3.11}, where the domain of integration
$ \R^3 \times S^2 $ is replaced by  $ \R^d \times S^{d-1} $.
Hence, Proposition \ref{Prop3.1} can be formally  
generalized to the case of arbitrary dimension $ d \geq 2 $. For simplicity of presentation we shall consider below
mainly the case $ d=3 $.

We present without derivation two other useful forms of
integral~\eqref{eq-3.4} \cite{BK}. The first form reads
\begin{multline}
		K[f](v) = \frac18 \int \limits_{\R^3 \times \R^3} dw\, dk\, \delta (k \cdot u) R(u - k/2, u + k/2) \times\\
		\times F[(f(v),f( w + k/2); f(v + k / 2), f(w)],\quad u = v - w.
		\label{eq-3.14}
	\end{multline}
It is clear that the integral over $ k \in \R^3  $ can
be reduced to the integral over the plane orthogonal to
$ u \in \R^3 $. The transformation of that kind was firstly used for the Boltzmann equation by Carleman \cite{Car}. The
second form of $ K[f](v) $ reads 
\begin{equation}
		K[f](v) = \frac14 \int \limits_{\R^3 \times S^2} dw\, dn\, |u \cdot n| R(u, u') F[f(v),f(w); f(v'), f(w')],\label{eq-3.15} 
		\end{equation}
		where
		\begin{equation}
\ u = v - w, \quad	n \in S^{2}, \quad	v' = v - (u \cdot n) n,\quad w' = w + (u \cdot n) n,\quad u' = v' - w'.\label{eq-3.16} 
	\end{equation}
Note that the notations for $ v' $ and $ w' $ in this
formula differ from similar notations in \eqref{eq-3.9}.
We shall not use below the representation \eqref{eq-3.15},\eqref{eq-3.16}
of $  K[f](v)$, but it should be mentioned because in
the case \eqref{eq-3.5} of the \BE it is the most 
conventional form of the collision integral (see e.g.
books \cite{Cer1}, \cite{CIP}). If the kernel $ R(u_1, u_2) $ is isotropic we can use equality  \eqref{eq-3.13} and
obtain
\begin{gather}
		K[f](v) = \int \limits_{\R^3 \times S^2} dw\, dn\,{B}(|u|, \hat{u} \cdot n)\, F[f(v),f(w); f(v'), f(w')],
	\label{eq-3.17}
		\\
		\intertext{in the notation of \eqref{eq-3.16},
		where}
		{B}(|u|, \hat{u} \cdot n) = 2|\hat{u} \cdot n|\,
		g[|u|,1 - 2(\hat{u} \cdot n)^2],\qquad
		\hat{u} = u / |u|.\nonumber
	\end{gather}
	
In the next section we discuss some basic properties of
kinetic equation \eqref{eq-3.3}, \eqref{eq-3.4}.

\subsection{Conservation laws and generalized $ H $-theorem}

We denote for any test function $ h(v) $
\begin{equation}\label{eq-3.18} 
		\langle f, h \rangle = \int \limits_{\R^3} dv\, f(v) h(v)
	\end{equation}
assuming that  the integral exists. If $ f(v,t) $ is 
a solution of \eqref{eq-3.3} we formally obtain
\begin{equation}\label{eq-3.19}  
		\frac{d}{dt}\langle f, h \rangle = \langle K[f], h \rangle.
	\end{equation}
 After straightforward 
transformations by using $  K[f](v) $, for example,  in the form  \eqref{eq-3.11}, we obtain
\begin{gather}
		\langle K[f], h \rangle = -\frac14 \int \limits_{\R^3 \times \R^3 \times S^2} dv\, dw\, d\omega
		\, {g}(|u|, \hat{u} \cdot \omega) \;  G(v, w; v', w')\times \nonumber\\
\qquad \qquad		\times [h(v') + h(w') - h(v) - h(w)], \label{eq-3.20} \\
\intertext{where}
		G(v,w; v', w') = F[f(v), f(w); f(v'), f(w')],\label{eq-3.21} 
	\end{gather}
for any function $ F(x_1,x_2; x_3,x_4) $ satisfying
conditions \eqref{eq-3.1}. By considering the functional 
equation
\begin{equation}\label{eq-3.22} 
 h(v') + h(w') - h(v) - h(w) = 0
\end{equation}	
in the notation of 	\eqref{eq-3.9} one can easily check that two scalar functions $ h_1=1 $, $  h_3 =|v|^2  $
and also the vector-function $ h_2 = v \in {\R^3}$	are
linearly independent solutions of this equation. The
uniqueness of these solutions in different classes of
functions is proved by many authors (see the discussion on equation (3.5) in the book \cite{CIP})

	Hence, we have the following conservation laws for equation \eqref{eq-3.3}, \eqref{eq-3.4}:
	\begin{equation}\label{eq-3.23}
		\langle f, 1 \rangle = \text{const},\ \langle f, v \rangle = \text{const},\ \langle f, |v|^2 \rangle = \text{const},
	\end{equation}
	provided the conditions \eqref{eq-3.1} for $ F $ in \eqref{eq-3.4} are fulfilled. The corresponding integrals in \eqref{eq-3.23} in the case \eqref{eq-3.5}  of the \BE have respectively physical meaning of total number of particles (gas molecules), total momentum and total kinetic energy.  These properties of the \BE  were discussed above in Section 2.6.

	Let us assume that there exists a function $   p(x) $
such that
	\begin{equation}\label{eq-3.24}	
	F(x_1,x_2; x_3,x_4) \, [ p(x_3)  +  p(x_4)  -  p(x_1) 
	- p(x_2) ]  \geq 0
\end{equation}	
for almost all $ x_i \geq 0, i=1,2,3,4 $. Then we can
formally introduce a generalized $ H $-functional (see
the end of Section 2.6) on a set of non-negative solutions 
$ f(v,t) $ of equation \eqref{eq-3.3}, \eqref{eq-3.4} by
formula 
\begin{equation}\label{eq-3.25}	
	\hat{H} [f(\cdot,t)] = \int\limits_{{\R^3}} dv I[f(v,t)],
	\quad
	I(x)= \int\limits_{0}^{x} dy p(y),
\end{equation}	
assuming the convergence of integrals. Then formal 
differentiation yields
\begin{equation*}	
	\frac{d}{dt} \hat{H} [f(\cdot,t)] = 
\langle f_t, p(f) \rangle 	= \langle K(f), p(f)\rangle. 
\end{equation*}	
	We always assume that $ F $ in \eqref{eq-3.4} satisfies
	conditions  \eqref{eq-3.1}. Therefore we can apply
the identity \eqref{eq-3.20} and 	 conclude by inequality
\eqref{eq-3.24} that $ \hat{H} [f(\cdot,t)]  $ cannot increase in time. In case \eqref{eq-3.5} of the \BE the
inequality 	\eqref{eq-3.24} holds for $ p(x) = \log x $
	and we obtain
\begin{equation}\label{eq-3.26}
\hat{H}(f) = \langle f(v,t),\; \log f(v,t) -1 \rangle.	
\end{equation}	
Note that $  \langle f,1\rangle= const.$  because of
conservation laws \eqref{1.1}. Therefore the functional
$\hat{H}(f)   $  is basically the same  as the classical
Boltzmann's $ H $-functional $ H(f)= \langle f, \log f\rangle $ considered in Section 2.6.
	
The results of this section can be formulated as follows.

\begin{Prop}\label{Prop3.2}	
The equation  \eqref{eq-3.3}, \eqref{eq-3.11}, where 
$ F(x_1,x_2; x_3,x_4) $ satisfies conditions  \eqref{eq-3.1}, has the same conservation laws 
\eqref{eq-3.23}, as the \BE. If the function $ F $ also
satisfies inequality \eqref{eq-3.24} for some function
$ p(x) $, $  x \in \R_+ $, then  (at least formally) the
functional $  \hat{H} [f(\cdot,t)$  \eqref{eq-3.25} on any
solution $ f(v,t) $ of equation \eqref{eq-3.3}, \eqref{eq-3.11} 
cannot increase in time $ t \geq 0 $.

\end{Prop}

	 Of course, all our considerations in Sections 3.1, 3.2
	 were done at the formal level of mathematical rigour, since we did not specify the function
 $ F(x_1,x_2; x_3,x_4) $ in \eqref{eq-3.4}. In the next section we consider some specific cases, which are different from the Boltzmann case \eqref{eq-3.5}, but also
 interesting for applications.

 \subsection{Nordheim--Uehling--Uhlenbeck equation and wave \\ kineticequation}
 
 It is clear that specific operators $ K $ from
 \eqref{eq-3.4} can have different functions
$ F(x_1,x_2; x_3,x_4) $. In all interesting  applications 
the function  $ F $ can be represented as a difference of two functions
\begin{equation}\label{eq-3.27}
F(x_1,x_2 ; x_3,x_4) = P(x_3,x_4 ; x_1,x_2) - P(x_1,x_2 ; x_3,x_4).
\end{equation}
 There are at least three cases of kinetic equations of interest to physics \eqref{eq-3.3}, \eqref{eq-3.4}, for
which    $ F $  has the structure \eqref{eq-3.4} with
different functions $ P $. These are the following cases:\\
(A) Classical Boltzmann kinetic equation
	\begin{equation}\label{eq-3.28}
		P_B(x_1,x_2 ; x_3,x_4) = x_1 x_2;
	\end{equation}
	(B) Quantum Nordheim--Uehling--Uhlenbeck equation  for bosons and  fermions \cite{Nor, UU}
	\begin{equation}\label{eq-3.29}
		P_{NUU}(x_1,x_2 ; x_3,x_4) = x_1 x_2 (1 + \theta x_3)(1 + \theta x_4),
	\end{equation}
	where $ \theta = \pm 1; $
\\
	(C) Wave kinetic equation (WKE) (see \cite{EV}, \cite{DymK}
	 and references therein)
	\begin{equation}\label{eq-3.30}
		P_W(x_1,x_2 ; x_3,x_4) = x_1 x_2 (x_3 + x_4).
	\end{equation}

	By using similar notations for $ F $ it is easy to verify that
	$$ F_{NUU}(x_1,x_2 ; x_3,x_4) = F_B(x_1,x_2 ; x_3,x_4) + \theta F_W(x_1,x_2 ; x_3,x_4), $$
because $ \theta^2 = 1 $ and therefore terms of the 
fourth order in $ F_{NUU}  $ vanish. A review of
mathematical results for NUU-equation can be found in
\cite{Ar3}, \cite{Ar4}. An interesting formal generalization of this equation to the case of so-called
anions (quasi-particles with any fractional spin between 0 and 1) is also considered in these papers. This model 
corresponds to \eqref{eq-3.27}, where\\
(D)
\begin{gather}\label{eq-3.31}
 P(x_1,x_2 ; x_3,x_4) = x_1 x_2 \Phi(x_3) \Phi(x_4),\\
		\nonumber
		\Phi(x) = (1 - \alpha x)^\alpha [1 + (1 - \alpha) x]^{1-\alpha},\quad 0 < \alpha < 1.
	\end{gather}

The limiting values $ \alpha =0,1 $ correspond to
NUU-equation for bosons ($ \alpha =0 $) and fermions 
($ \alpha = 1$). The existence of global solutions in
$ L^1\cap L^\infty $ of the Cauchy problem for equation
  \eqref{eq-3.3}, \eqref{eq-3.4}, where $ F $ is given in
\eqref{eq-3.27}, \eqref{eq-3.31} is proved in 
\cite{Ar3}
under some restriction on initial conditions and the
kernel $ R $ of the operator  \eqref{eq-3.4}.

We note that in all above cases (A), (B), (C), (D) it is
possible to find  the function $ p(x) $ that  satisfies
inequality \eqref{eq-3.24}. Indeed the function $ F $
for cases (A), (B) and (D) can be written as
\begin{gather}\label{eq-3.32}
F(x_1,x_2 ; x_3,x_4) =x_3  x_4 \Phi(x_1) \Phi(x_2)
 - x_1 x_2 \Phi(x_3) \Phi(x_4) =
\\
\nonumber
= 
[\Psi(x_3) \Psi(x_4) -  \Psi(x_1) \Psi(x_2)]
\prod\limits_{i=1}^{4} \Phi(x_i),\quad
\Psi(x) = \frac{x}{\Phi(x)},
\end{gather}
where  $   \Phi(x) = 1  $  for the case (A),
$  \Phi(x) =  (1+ \theta x)^{-1} $ for the case (B),
and $   \Phi(x)   $  is given  in \eqref{eq-3.31} for
the case (D). Then it is easy to see that
\[ p(x) = \log \Psi(x) = \log x - \log \Phi(x) \]
satisfies \eqref{eq-3.24} provided $ \Phi(x) > 0 $.
The positivity condition for $ \Phi(x)  $ is fulfilled for all
$ x \geq 0 $ in the case (A) and (B) with $ \theta =1 $.
It is also fulfilled for $ 0 \leq x  < 1 $ in the case
(B) with $  \theta = -1 $ and for $  0 \leq x  < 1/\alpha $ in the case (D). The known  results on existence of 
solutions of kinetic equations \eqref{eq-3.3} with
corresponding operators \eqref{eq-3.4} show that it is
sufficient to satisfy these restrictions   only at 
$ t=0 $  \cite{Ar3}. Thus the kinetic equation \eqref{eq-3.3} has in cases (A), (B), (D) the monotone
decreasing functionals
\[	
\hat{H}	[f(\cdot, t) ] =\int \limits_{\R^3}  dv
I[f(v,t)]\quad
I(x) = \int \limits_{0}^{x}  dy [\log y - \log\Phi(y) ] ,
\]
with corresponding functions 	$ \Phi(x) $.

The inequality \eqref{eq-3.24} in cases (A), (B) and (D)
allows to solve an important question about stationary solutions of equation \eqref{eq-3.3}. If $ K[f^{st}](v)=0 $,
then we can integrate this equality against any "nice" function $ h(v)  $ and obtain the identity (with respect
to $ h(v)  $)  $  \langle   K(f^{st}), h\rangle = 0$, in
the notation of \eqref{eq-3.18}. Then we take  
$ h(v)  =\log \Psi[f^{st}   ]   $ and use the transformation \eqref{eq-3.20} and inequality \eqref{eq-3.24}.
Since the resulting integral of non-negative function over
the set $ \R^3 \times \R^3  \times S^2   $	must be equal
to zero, we conclude that this function is equal  to zero
almost everywhere in that set. This leads to equation
\begin{equation*}
h(v') +h(w') -h(v) -h(w)= 0, \quad
h(v)= \log \Psi[f^{st}   ],
\end{equation*}	
in the notation of 	\eqref{eq-3.9}, \eqref{eq-3.32}. The 
equation holds almost everywhere in
$ \R^3 \times \R^3  \times S^2   $. Then we obtain (see 
comments to equation 	\eqref{eq-3.22} above)
\begin{equation*}
\Psi[f^{st}(v)   ] = \frac{f^{st}(v)}{\Phi[f^{st}(v)]}
= M(v)= \exp (\alpha +  \beta \cdot v + \gamma |v|^2 ),
\end{equation*}	
where $ \alpha \in \R $, $ \gamma \in \R $,
$ \beta \in \R^3 $ are arbitrary  constant parameters.
 For brevity we do not discuss these known stationary
 solutions, see e.g. \cite{Ar4} for
 details. We stress
 that the above considerations just repeat usual arguments
  in the proof of uniqueness of the Maxwellian stationary solution to the \BE, see e.g. \cite{Cer1}. Because of 
many similarities with the Boltzmann case (A) one can 
expect similar behaviour of solutions to equation \eqref{eq-3.3}  for cases (B), (D), in particular, convergence
to above discussed stationary solutions for large values
of time.

The situation looks more complex in the case  (C) of WKE.
In that case we can also satisfy the inequality \eqref{eq-3.24} by choosing $ p(x) = -1/x $. Then this
inequality with $ F $ from \eqref{eq-3.27}, \eqref{eq-3.30} reads
\begin{equation}\label{eq-3.33}
x_1 x_2 x_3 x_4 ( x_1^{-1} + x_2 ^{-1} -x_3^{-1}
-x_4^{-1})^2  \geq 0.
\end{equation}
However, the attempt to construct the  $ \hat{H} $-functional 	\eqref{eq-3.25} leads to divergent integral
$ I(x) $. It looks reasonable to replace this integral in
\eqref{eq-3.25} to $  I(x) = - \log x   $, then we formally 
obtain
\[
\hat{H}[f(\cdot, t) ] = - \int \limits_{\R^3 } dv \log
f(v,t).
\]
This integral is divergent for large $ |v| $, because
we always assume that $  f(v,t) \to 0 $, as
 $ |v| \to \infty $. Below we shall try to clarify the
 situation with WKE by using discrete kinetic models introduced in the next section.

	\subsection{Discrete kinetic models}
The idea of using discrete velocity models for qualitative 
description of solutions to the \BE seems to be very natural.
 Implicitly it was already 	used by Boltzmann in the first 
 publication \cite{Bol} of his famous equation. We also mention
 first two toy-models with a few velocities introduced  by
Carleman \cite{Car}  and Broadwell \cite{Bro} respectively.
An important role in the development of this idea was played 
by Cabannes \cite{Cab} and Gatignol \cite{Gat} in 1980s. 
Moreover, it was proved in 1990s (see \cite{BPS}, \cite{PSB}
and references therein) that the \BE can be approximated by its
discrete velocity models when the number of velocities tends
to infinity. These results show that discrete models can be used not only for  qualitative, but also for quantitative 
description of solutions to the \BE.

The similar scheme of construction of discrete models can be
applied to the general kinetic equation \eqref{eq-3.3}, \eqref{eq-3.4}. We introduce the velocity space $ V \subset \R^d $
that contains $ n \geq 4 $	points and replace the function
$ f(v,t) $  by a vector  $ f(t) \in \R^n $, where  
\begin{equation}\label{eq-3.34}
V = \{ v_1, \dots,v_n  \},
\;\quad  
f(t) = \{ f_1(t),\dots,f_n(t) \}.
\end{equation}	
It is implicitly assumed here that 	$ f_i(t) $ approximates
for large $ n $ the function $ f(v,t) $  at the point
$ v= v_i \in \R^d, \;i=1, \dots, n  $.  Speaking  about
discrete models it  is convenient to use an arbitrary dimension
$ d \geq 2 $, as we shall see below. The simplest and the most
transparent case is, of course, the plane case $ d=2 $. The
kinetic equation \eqref{eq-3.3}, \eqref{eq-3.4} in the $ d $-dimensional
case \eqref{eq-3.12} is replaced by the following set of ordinary
 differential equations
\begin{equation}\label{eq-3.35}
	\frac{df_i}{dt} = \sum_{j,k,l=1}^{n}
	 \G_{ij}^{kl} F_w(f_i,f_j ; f_k,f_l),
	\quad
	\G_{ij}^{kl} = \G_{ji}^{kl} = \G^{ij}_{kl},
	\quad
	1 \leq i \leq n,
\end{equation} 	
where the constant (for given set $ V $) parameters $ \G_{kl}^{ij}$  depend only on $ |v_i - v_j | =   |v_k - v_l | $ and
$ (v_i- v_j) \cdot (v_k- v_l) $ for any  integer values of indices $  1 \leq i, j, k, l \leq n $. The strict inequality
$ \G^{kl}_{ij}  >  0 $	is possible only if
\begin{equation}\label{eq-3.36}	
	v_i + v_j = v_k + v_l,\quad |v_i|^2 + |v_j|^2 = |v_k|^2 + |v_l|^2.
\end{equation} 	
Note that equations \eqref{eq-3.35}	 have a universal form for any dimension $  d \geq 2 $, though the coefficients $\G^{kl}_{ij}   $ can depend on $ d $. The equalities \eqref{eq-3.36}
have a simple geometrical meaning: the points $ \{v_i, v_j,
v_k, v_l \} $ form a  rectangle, where the two pairs of points 	
$ \{v_i, v_j \} $	and $ \{v_k, v_l \} $	 belong to two 
different diagonals.
%
%
%
%
%
%
%
%
%
	
Obviously, this geometrical meaning does not depend on dimension.  The simplest non-trivial example of the
set $ V \subset \R^2 $	from \eqref{eq-3.34} has just four "velocities" 
\begin{gather}
	v_1 = (1,0),\ v_2 = (-1,0),\ v_3 = (0,1),\ v_4 = (0,-1);\nonumber\\
	f(v_i,t) = f_i(t),\ i = 1, \dots, 4; 
	\quad \G_{12}^{34} = \G_{21}^{34} = \G_{34}^{12} = 1. \label{eq-3.37}
\end{gather}	
like in the plane Broadwell model \cite{Bro}	of the \BE.
Equations \eqref{eq-3.35} of the model read as
\begin{equation}\label{eq-3.38}
	 \frac{\partial f_1}{\partial t} = \frac{\partial f_2}{\partial t} = -\frac{\partial f_3}{\partial t} = -\frac{\partial f_4}{\partial t} = \G^{34}_{12} F(f_1,f_2 ; f_3,f_4),
\end{equation}	
where  $  \G^{34}_{12} > 0 $ is a constant. This equation for
the Boltzmann case 	\eqref{eq-3.25} can be easily reduced to
 the linear equation. The case \eqref{eq-3.27}, \eqref{eq-3.30}  of WKE is a bit
 more complicated, it is discussed in \cite{BK} in more detail.

 In order to construct a discrete model \eqref{eq-3.34}, \eqref{eq-3.35}, which has all relevant properties of the
 initial kinetic equation \eqref{eq-3.3}, \eqref{eq-3.4}, we
 need to impose some restrictions on the set $ V $ and the
 coefficients of equations  \eqref{eq-3.35}.
 
\noindent{\it{Definition 1}  The model  \eqref{eq-3.34},  \eqref{eq-3.35}  is called \underline{normal}, if it satisfies
the following conditions on the set $ V $: (a) all its $ n $
elements are pairwise different and do not lie in a linear
subspace of dimension $  d' \leq d-1 $	or on the sphere in
$ \R^d $; (b) the set $ V $ does not have isolated points, i.e.
for any $ 1 \leq i \leq n $ in \eqref{eq-3.35} there exist such
$ 1 \leq j, k, l \leq n $ that $ \G_{ij}^{kl} > 0$;
(c) if the functional equation
\begin{equation}\label{eq-3.39}	
	h(v_i) + h(v_j) -h(v_k) - h(v_l) = 0
\end{equation}		
is fulfilled for \underline{all} indices $ (i,j; k,l) $	 for
which $ \G_{ij}^{kl}  > 0 $, then there exist such constants 
$ \alpha, \gamma \in \R  $, $ \beta \in \R^d $ that	
	$ h(v) = \alpha +  \beta \cdot v + \gamma |v|^2  $.	}	

The methods of construction of normal models are discussed in
more detail in Appendix A. All models that we consider below are
assumed to be normal.

\subsection{Properties of discrete models}
	
The discrete kinetic models 	\eqref{eq-3.34}, \eqref{eq-3.35} of the kinetic equation \eqref{eq-3.3}, \eqref{eq-3.4} is uniquely defined by (a) the function 
$ F(x_1, x_2; x_3,x_4) $, satisfying conditions \eqref{eq-3.1};
(b) the phase 	 $   V = \{v_1, \dots, v_n  \} \subset \R^d $;
(c) the set  of coefficients	
$\G = \{ \G_{ij}^{kl}  \geq  0, 1 \leq i, j, k, l \leq n \}$,	where  $ \G_{ij}^{kl}  $  can depend on  $ |v_i - v_j | =   |v_k - v_l | $ 	and $ (v_i - v_j) \cdot (v_k - v_l)   $.
We remind that the inequality 	  $ \G_{ij}^{kl}  > 0 $ is
possible only under conditions \eqref{eq-3.36}. Moreover the 
symmetry conditions from \eqref{eq-3.35}  are fulfilled for all
elements of the set $ \G $.

By using these symmetry conditions for  $ \G_{ij}^{kl} $ and
related conditions \eqref{eq-3.1} for $ F $ it is easy to 
derive from \eqref{eq-3.35} the following identity
\begin{equation}\label{eq-3.40}
\frac{d}{dt} \sum \limits_{i=1}^{n} f_i(t) h_i =
- \frac{1}{4} \sum \limits_{i,j,k,l = 1 }^{n} \G_{ij}^{kl}
F(f_i, f_j; f_k, f_l) (h_k+ h_l -h_i-h_j ),
\end{equation}	
where $ h_1, h_2, \dots, h_n $	 are constant numbers.
Obviously, this is a discrete analogue of the identity
\eqref{eq-3.20} for kinetic equation
\eqref{eq-3.3}, \eqref{eq-3.4}. Then the conditions \eqref{eq-3.36}  lead to following conservation laws
\begin{equation}\label{eq-3.41}	
\sum \limits_{i=1}^{n} f_i(t) = const.,
\quad
\sum \limits_{i=1}^{n} f_i(t)v_i = const.,
\quad
\sum \limits_{i=1}^{n} f_i(t)  |v_i|^2 = const.
\end{equation}	
similar to integrals  \eqref{eq-3.23} for the kinetic equation.
Note that both the identity 	\eqref{eq-3.40}  and conservation laws
\eqref{eq-3.41}  are valid for any discrete kinetic model, not
only for normal models which  cannot have other linear conservation
laws than the ones listed in \eqref{eq-3.41}  or their linear
combinations. This is, however, true not for any normal model,
but at least for normal models with function $ F $ satisfying
the inequality \eqref{eq-3.24}  for some function $ p(x) $,  $ x > 0 $. We can prove the following statement.
\newtheorem{Theorem}{Theorem}[section]
\begin{Theorem}\label{Th3.1}
Assume that the model \eqref{eq-3.34},  \eqref{eq-3.35} is
normal and there exists  a function $  p(x)  $ such that
the inequality  \eqref{eq-3.24} is satisfied for all 
$ x_i > 0, i=1,2,3,4 $. Assume also that
\begin{itemize}
\item[(1)] there exist two numbers $ 0 < a < b $ such  that
$  p(x)  $ is continuous and strictly  monotone for all
$ x \in [a,b] $;
\item[(2)] the equality sign in \eqref{eq-3.24} is possible
only if 
\[  p(x_3)  +p(x_4) - p(x_1) - p(x_2) =0.
\]
\end{itemize}
Then 
\begin{itemize}
\item[(a)]
there exists a function $ H(x_1,\dots,x_n) $   such that
\begin{equation}\label{eq-3.42}	
\frac{d}{dt}  H[f_1(t), \dots, f_n(t) ] \leq 0
\end{equation}
for any solution   $ \{ f_i(t)  > 0,   \quad i=1,\dots,n  \} $
of \eqref{eq-3.35};  
$      $
\item[(b)]
if
\begin{equation}\label{eq-3.43}	
\frac{d}{dt} \sum \limits_{i=1}^{n} f_i(t) h_i = 0
\end{equation}
for any solution    $ \{ f_i(t)  > 0,    i=1,\dots,n  \} $
   of \eqref{eq-3.35}, then
$ h_i = \alpha +  \beta \cdot v_i + \gamma |v_i|^2  $ in the
notation of \eqref{eq-3.34}, where 
$ \alpha \in  \R $, $ \gamma \in  \R $ and $ \beta \in  \R^d $
are some constant parameters;
\item[(c)]
if  $ f^{st} = \{ f^{st}_{1}, \dots, f^{st}_{n}  \} $ is 
a stationary solution of \eqref{eq-3.35}, then
\begin{equation}\label{eq-3.44}	
p( f^{st}_{i} )  =  \alpha +  \beta \cdot v_i + \gamma |v_i|^2,
\quad
i=1,\dots,n,
\end{equation}
for some constant parameters $ \alpha \in  \R $, $ \gamma \in  \R $ and $ \beta \in  \R^d $.
\end{itemize}
\end{Theorem}
 \begin{Proof}
Let $ I(x) $ be any function such that $ I'(x) = p(x) $  for
all $  x > 0 $. Then we denote 
\begin{equation}\label{eq-3.45}
 H(x_1, \dots,x_n )
= \sum \limits_{i=1}^{n} I(x_i).
 \end{equation}
 If $ \{ f_i(t) \;    i=1,\dots,n  \} $ satisfy \eqref{eq-3.35}
 then 
 \begin{gather}\label{eq-3.46}
 \frac{d}{dt}  H[f_1(t), \dots, f_n(t) ] =
 \sum \limits_{i=1}^{n} p(f_i) \frac{d f_i}{dt} =
\\
\nonumber
= - \frac{1}{4} \sum \limits_{i,j,k,l = 1 }^{n} \G_{ij}^{kl}
F(f_i, f_j; f_k, f_l) 
[p(f_k)+ p(f_l) -  p(f_i)  - p(f_j) ] \leq 0
 \end{gather}
 as it follows from \eqref{eq-3.40}, \eqref{eq-3.24}. Hence,
 (a) is proved.

 To prove point (b) we use  \eqref{eq-3.46} and reduce 
\eqref{eq-3.42} to identity
 \begin{equation}\label{eq-3.47}
 \sum \limits_{i,j,k,l = 1 }^{n} \G_{ij}^{kl}
F(f_i, f_j; f_k, f_l) (h_k+ h_l -h_i-h_j ) = 0,
 \end{equation}
which is supposed to be valid for any $  f_i > 0 $, \;
 $  i=1,\dots, n  $. Let us assume that (b) is wrong and that
$  h = (h_1,  \dots, h_n ) $  in this identity is not a linear
combination of vectors $\varphi_1,  \dots, \varphi_{d+2}   $ in
the notation of (A2) from Appendix A. Without loss of generality
we can assume that
\begin{equation}\label{eq-3.48}
\alpha \leq h_i  \leq \beta, \quad   i=1,\dots, n, 
\end{equation}
 where $   \alpha <  \beta $ is any pair of given real numbers.
 Indeed if $  h = (h_1,  \dots, h_n ) $ satisfies \eqref{eq-3.47},
  then so does
 \begin{equation*}
 \tilde{h}= \lambda h + \mu \varphi_1,
 \quad
 \varphi_1= (1,1,\dots, 1),
  \end{equation*} 
  where $ \lambda $ and $ \mu $ are any real numbers. We can
  always choose these numbers in such a way that conditions
 \eqref{eq-3.48} for $ \tilde{h} = ( \tilde{h}_1, \dots, \tilde{h}_n ) $ are fulfilled. Tildes are omitted below. The
 numbers  $  \alpha $ and $ \beta $ in
 \eqref{eq-3.48} are chosen in the following way. It follows
 from assumption (1) of the theorem that the function $ p(x) $
 maps  the interval $ [a,b] $ to some other interval, say,
 $ [\alpha, \beta ]$. Moreover there is an inverse function
 $ x(p) $, which maps any point $ p \in [\alpha, \beta ]$ to
 $ x(p) \in   [a,b] $. Then we can easily construct a 
 counterexample to our  assumption by substituting
 $ f_i = x (h_i), \quad  i=1,\dots, n $, into identity 
\eqref{eq-3.47}. We obtain a sum of non-negative terms and
conclude that each term vanishes, i.e. 
$ F[ x(h_i), x(h_j); x(h_k), x(h_l)   ](h_k + h_l - h_i - h_j) = 0$   
for any $ 1 \leq i,j,k,l \leq n $ such that
$\G_{ij}^{kl} > 0 $.  Then we use assumption (2) of the theorem 
and   Definition 4.1. This proves (b).

 In order to prove (c) we consider identity \eqref{eq-3.47} for
 a stationary solution $ f^{st} $. Then the left hand side
 of \eqref{eq-3.47} is equal to zero for arbitrary vector 
 $ h = (h_1, \dots, h_n) $. We substitute $ f_i = f^{st}_i $,
 $ h_i = p(f^{st}_i  ), \quad  i=1,\dots, n $, into  \eqref{eq-3.47} and obtain again a sum of non-negative terms. The same considerations as above lead to equalities
$$ p( f^{st}_k) +  p( f^{st}_l) - p( f^{st}_i) -
p( f^{st}_j) = 0 $$
for all  $ 1 \leq i,j,k,l \leq n $ such that $\G_{ij}^{kl} > 0 $.
Then we again use Definition 4.1 and prove (c). This completes
the proof.
\end{Proof}

	\subsection{Some transformations of equations and initial data}
We return for a moment to kinetic equation 	\eqref{eq-3.3},
\eqref{eq-3.4} and note that this equation is invariant under
rotations and translations of variable $ v \in \R^3 $ (or
 $ v \in \R^d, \; d \geq 2 $, in the general case, see \eqref{eq-3.12}). Usually we consider such initial data 
$ f(v,0) \geq 0  $ for \eqref{eq-3.3} that 
	\begin{equation}\label{eq-3.49}
	\int \limits_{\R^3}  dv (1+ |v|^2) f(v,0)  < \infty.
\end{equation}	
	
Invariance of equation 	\eqref{eq-3.3},
\eqref{eq-3.4}  under translations means that if $ f(v,t) $ is
a solution of this equation, then so is  $ f_a(v,t) = f(v+a,t)   $ for any  $ a \in \R^3 $. Then
		\begin{equation}\label{eq-3.50}
	\langle  f_a(v,0), v \rangle  = 
\langle  f(v,0), v \rangle - a  \langle  f(v,0), 1 \rangle 
\end{equation}	
in the notation of 	 \eqref{eq-3.18}. It is always assumed that
$  \langle  f(v,0), 1 \rangle \neq 0   $.  Therefore we can
always choose $ a \in \R^3 $ ( or $ a \in \R^d $ in the
general case)  in such a way that $ \langle  f_a(v,0), v \rangle  = 0 $. This almost trivial  observation allows to
consider only such initial conditions for \eqref{eq-3.3},
\eqref{eq-3.4}	 that $  \langle  f(v,0), v \rangle  = 0 $.

This transition from general kinetic equation  \eqref{eq-3.3}, \eqref{eq-3.4}	to its  discrete model  \eqref{eq-3.34}, \eqref{eq-3.35}	preserves the translational symmetry of
equations \eqref{eq-3.35}. Indeed the shifting $ v \to v+a $
of $ v $-variables means for the discrete model the transformation of the set $ V $	 from \eqref{eq-3.34}
\begin{equation}\label{eq-3.51}
V = \{v_1, \dots, v_n  \} \rightarrow
V = \{v_1+a, \dots, v_n +a \}, \quad	a \in \R^d. 
\end{equation}

Equations \eqref{eq-3.35}	 of the model are connected with the
set $ V \subset \R^d $ only through  coefficients
 $\G_{ij}^{kl} $. However these coefficients  depend  (for
 fixed indices $ 1 \leq i,j,k,l \ n $) only on differences
 $( v_i - v_j)$  and $( v_k- v_l) $.
 Therefore each coefficient $\G_{ij}^{kl} $  is invariant 
 under translations \eqref{eq-3.51}	 of the whole set $ V $.
 It is also easy to check that the Definition 4.1 of normal 
 discrete model is invariant under translation. In order words,
 if the pair $ (V, \G) $, where 
 $ \G = \{\G_{ij}^{kl}, \; 1 \leq i,j,k,l  \} $, defines a
 normal model, then so does any pair $ (V_a, \G) $ in the
 notation of \eqref{eq-3.51}.

\section{Convergence to equilibrium for discrete models\\ of wave kinetic equations}	 

\subsection{Statement of the problem and formulation of results}	 
We consider below the discrete models 
	  \eqref{eq-3.34}, \eqref{eq-3.35}, where	
\begin{equation}\label{eq-4.1}
F(x_1,x_2; x_3,x_4) = x_3 x_4 (x_1+x_2) - x_1 x_2 (x_3 + x_4),
\end{equation}	
i.e. the models of WKE \eqref{eq-3.3},	\eqref{eq-3.4} with	
function $ F $ given in \eqref{eq-3.27},	\eqref{eq-3.30}.
The set of ODEs \eqref{eq-3.35} in simplified notations reads
\begin{equation}\label{eq-4.2}
\frac{df}{dt} = Q(f) = Q^{+}(f) - Q^{-}(f),
\end{equation}	
where 
\begin{gather}
\nonumber
f(t) = \{ f_1(t), \dots, f_n(t) \}, \; 
Q^{\pm}(f) = \{ Q^{\pm}_1(f), \dots, Q^{\pm}_n(f)   \},
\\
Q^{+}(f) = \sum \limits_{i,j,k,l = 1 }^{n} \G_{ij}^{kl} 
f_k f_l (f_i + f_j),
\quad 
Q^{-}(f) =  f_i B_i(f),\label{eq-4.3}
\\
B_i(f) =  \sum \limits_{j,k,l = 1 }^{n} \G_{ij}^{kl} 
f_j (f_k + f_l), \quad i=1,\dots, n; \quad 
\G_{ij}^{kl} =
\G_{ji}^{kl} =
\G_{kl}^{ij}, \quad
 1 \leq i,j,k,l \leq n.
 \nonumber 
\end{gather}
The constant coefficients $ \G_{ij}^{kl} \geq 0 $ depend on the
phase set 
\begin{equation}\label{eq-4.4}	
	V = \{v_i \in \R^d,  i=1,\dots, n\},
	\quad d \geq 2.
\end{equation}		
We remind that the 	strict inequality $ \G_{ij}^{kl} > 0 $
is possible only for such indices that $ v_i+v_j = v_k + v_l $,
\; $ | v_i|^2 + | v_j|^2 = | v_k|^2+| v_l|^2 $, 
$ 	1 \leq i,j,k,l \leq n  $. Each coefficient $ \G_{ij}^{kl} $
 depends on $ |v_i - v_j  |   =  |v_k - v_l  |  $ and
 $ (  v_i - v_j ) \cdot  ( v_k - v_l   )  $.

We consider the Cauchy problem for equations  \eqref{eq-4.2}
and initial conditions 
\begin{equation}\label{eq-4.5}	
f{|_{t=0}} = f^{(0)}= \{f_1^{(0)},\dots, f_n^{(0)}\},\;
   \quad
  f_i^{(0)} >0, \; i= 1, \dots, n; \quad  \sum_{i=1}^{n}f_i^{(0)} v_i =0.	
\end{equation}
The last restriction does not lead to a loss of generality, as
 explained in Section 3.6. A strict positivity condition is
  needed in order to avoid some "non typical" solutions and
  simplify proofs. For example, if $ \G_{12}^{34} > 0 $ and
  $ f_i^{(0)}=0  $ for all $  i \geq 5 $, then 
  $ f_5(t) = \dots = f_n(t) = 0 $ for $  t > 0 $. In that case
  we obtain from \eqref{eq-3.35} a simple set of four equations
\eqref{eq-3.38}, which should be considered separately
(see e.g. \cite{BK} ). Neglecting such special cases does not
 look very important for the general qualitative behaviour of
 solutions of the model.
 
 The main result of Section 4 can be formulated in the 
 following way.

\begin{Theorem}\label{Th4.1}
We assume that the discrete model	\eqref{eq-4.2}--\eqref{eq-4.4} is normal, i.e. the set $ V $ in \eqref{eq-4.4}
 and  the coefficients 
 $   \G_{ij}^{kl}, \;  1 \leq i,j,k,l \leq n $  satisfy
 Definition 1 from Section 3.4. It is  also assumed that
(1) $ v_1 = 0 $  in \eqref{eq-4.4} and  (2) if $ v_i \in V $,
then    $(-  v_i) \in V  $ for all  $ i=1,\dots, n $.
Then  the Cauchy  problem for equations \eqref{eq-4.2} and
the initial conditions \eqref{eq-4.5} has a unique solution
$ f(t) = \{ f_1(t), \dots f_n(t) \} $  for all $ t > 0 $.
Moreover, for all   $ 1 \leq i \leq n $,
 \begin{equation}\label{eq-4.6}
(a) \quad 0< f^{(0)}_{i}\exp(-c \rho_{0}^{2} t) \leq f_i(t) < \rho_0, \quad
\rho_0 =\sum_{i=1}^{n} f^{(0)}_{i},
\end{equation}  
where $ c > 0 $ is a constant independent of $  f^{(0)} $ ;

\begin{equation}\label{eq-4.7} 
(b) \lim_{t \to \infty} f_i(t) = a (1+ b |v_i|^2)^{-1},
\quad
 a \sum_{i=1}^{n} (1+ b |v_i|^2)^{-1}  = \rho_0,\vspace{-6pt}
\end{equation}  
where $  b > - M^{-1}$,
 $   M = \max \{ |v_i|^2, \; 1 \leq i \leq n\}$,
is a maximal real root of equation 
\begin{equation}\label{eq-4.8}
T_0   = \rho_0^{-1} \sum_{i=1}^{n} f^{(0)}_{i}|v_i|^2 =
\sum_{i=1}^{n} 
\frac{(1+ b |v_i|^2)^{-1} |v_i|^2}{\sum_{i=1}^{n}(1+ b |v_i|^2)^{-1}}.  
\end{equation}
 \end{Theorem}

  It is easy to see that the function $ T_0(b) $ defined by
  equality \eqref{eq-4.8} decreases monotonically on the
  interval $ - M^{-1}  \leq b < \infty $  from its maximal value
  $  T_0( -M^{-1})  = M  $ to zero for $ b \to \infty $.
  Therefore the root $ b(T_0) $  defined in the part (b) of
  the theorem is unique.
  
  The proof of Theorem \ref{Th4.1} is given in Sections 4.2--4.5.
   It is
based on simple estimates \eqref{eq-4.6}, conservation laws and
  on the fact that equations \eqref{eq-4.2} have a Lyapunov
  function which decreases monotonically on positive solutions
  of \eqref{eq-4.2}.

  \subsection{Existence and uniqueness of global non-negative solutions}
  
  We consider equations \eqref{eq-4.2}, \eqref{eq-4.3} and note that
\begin{equation}\label{eq-4.9}	
\sum \limits_{i = 1 }^{n}  Q_i(f)=
\sum \limits_{i = 1 }^{n}  Q_i(f) |v_i|^2	=0,
\quad
\sum \limits_{i = 1 }^{n}  Q_i(f) v_i =0,
\end{equation}	
for any $ f(t)= \{ f_1(t), \dots f_n(t) \} $. Therefore  any solution $ f(t) $	of the problem \eqref{eq-4.2}, \eqref{eq-4.5}
satisfies conservation laws

	\begin{equation}\label{eq-4.10}	
	\rho[  f(t) ] = 	\rho[  f_0 ] = \rho_0,
	\quad
	E[  f(t) ] = E(f_0) =  \rho_0 T_0,
	\quad
\sum \limits_{i = 1 }^{n}  f_i(t) v_i = 0,	
\end{equation}	
where
\begin{equation}\label{eq-4.11}	
	\rho(f) =   \sum \limits_{i = 1 }^{n}  f_i,
	\quad
	E(f) =  \sum \limits_{i = 1 }^{n}  f_i |v_i|^2.
\end{equation}	
The third equality  in \eqref{eq-4.10}  follows from \eqref{eq-4.5},  it will not be used in this section.

Note that the existence and uniqueness of local in time
solutions to the problem
\eqref{eq-4.2}, \eqref{eq-4.5} follow from general theory of ODEs. 
We, however, need to construct a global solution for positive
initial data \eqref{eq-4.5}.  In order to do it we use a
simple trick, which is more or less standard for the \BE 
(see e.g. \cite{CIP}). Namely, we modify equation 
\eqref{eq-4.2}  in the following way:
\begin{gather}
\nonumber
\frac{d \varphi}{dt},
 + \lambda \varphi = A(\varphi),
 \quad
 \varphi=(   \varphi_1, \dots,   \varphi_n),
 \\
 A(\varphi) = [ A_1(\varphi), \dots,   A_n(\varphi)],
 \quad
   A_i(\varphi) = Q_i(\varphi) + g  \varphi_i  \rho^2 (\varphi),
   \quad
   1 \leq i \leq n,
   \label{eq-4.12}	
\end{gather}	
where
\begin{equation}\label{eq-4.13}	
\varphi|_{t=0}  = f^{(0)},
\quad
\lambda = g \rho^2, \quad g = 2 \max \limits_{1 \leq i,j,k \leq {n} }
\sum \limits_{l = 1 }^{n} \G_{ij}^{kl}, 
\end{equation}	
in the notation of \eqref{eq-4.10}--\eqref{eq-4.11}.
Note that 	$ \lambda  $   and   $ g $ are positive constants. 
It is easy to see that  $   A_i(\varphi) \geq Q^+_i (\varphi ) \geq 0 $
because 
\begin{equation*}	
g \varphi_i \rho^2 (  \varphi)-  Q_i^-(\varphi )  =
\varphi_i  \sum \limits_{j,k = 1 }^{n} a_{jk}^{i}  \varphi_j  \varphi_k,
\end{equation*}	
where
\begin{equation*}
 a_{jk}^{i} = g - 2   \sum \limits_{l = 1 }^{n}
   \G_{ij}^{kl }\geq 0, \quad
1 \leq i,j,k \leq {n}. 
\end{equation*}	
Terms $ Q^{+}_{i}(\varphi)  $ also are polynomials with non-negative coefficients for any  $ 1 \leq i \leq n $.
Therefore, for any two vectors
\begin{equation}\label{eq-4.14}	
 \varphi=(   \varphi_1, \dots,   \varphi_n),
 \quad 
  \psi=(   \psi_1, \dots,   \psi_n)
\end{equation}	
with non-negative components $ \varphi_ i \geq 0 $  and 
$ \psi_ i \geq 0 $  and such that $ \varphi_ i \geq \psi_ i $
for all $ i =1, \dots, n $ we obtain
\begin{equation}\label{eq-4.15}	
 A_i(\varphi) \geq A_i(\psi) \geq  0 , \quad i=1,\dots, n.
\end{equation}

In order to construct the solution of the problem \eqref{eq-4.12}--\eqref{eq-4.13} we transform \eqref{eq-4.12}
into the integral equation
\begin{equation}\label{eq-4.16}	
 \varphi(t)  =  f^{(0)} e^{- \lambda t} +
 \int \limits_{0}^{t} d \tau  e^{- \lambda ( t- \tau)}
 A[  \varphi(\tau)  ]
\end{equation}	
and try to solve this equation by iterations
\begin{equation}\label{eq-4.17}	
 \varphi^{^{(k+1)}} (t)  =  f^{(0)} e^{- \lambda t} +
 \int \limits_{0}^{t} d \tau  e^{- \lambda ( t- \tau)}
 A[  \varphi^{(k)}(\tau)  ],
\end{equation}
$ k= 0,1, \dots $  and $ \varphi^{(0)} (t)  = 0 $.   Then 
it follows  from inequalities
\eqref{eq-4.15} that 
\begin{equation}\label{eq-4.18}	
0 < f^{(0)}_{i }  e^{- \lambda t} \leq \varphi^{(k)}_{i} (t)
\leq \varphi^{(k+1)}_{i} (t), \quad 1 \leq i \leq n,  \quad
 k= 1,2, \dots 
\end{equation}	
Hence, we obtain a monotone increasing sequence of positive
functions. It remains to prove that it is bounded above.

To this goal we consider a sequence of sums (with some abuse of
notation of \eqref{eq-4.11})
\begin{equation}\label{eq-4.19}	
\rho^{(k) }(t) = \sum \limits_{i= 1 }^{n} \varphi^{(k)}_{i} (t),
\quad
n= 0,1, \dots.
\end{equation}
By using the first identity from \eqref{eq-4.9}	 and the
definition of $  A(\varphi)  $ from \eqref{eq-4.12} we obtain
\begin{equation}\label{eq-4.20}	
\rho^{(k+1) }(t) =  \rho^{(0) }  e^{- \lambda t} +
 g \int \limits_{0}^{t} d \tau  e^{- \lambda ( t- \tau)}
[\rho^{(k) }(\tau)]^{3} , \quad
k \geq 0,  
\end{equation}
where
$ \lambda = g   \rho_{0 }^{2}$. By induction we can easily prove
that       $ \rho^{(k) } \leq    \rho_{0 }  $     for all        $ k \geq 0  $    because 
\begin{equation*}
 \rho_{0 }   e^{- \lambda t} + g  \rho_{0 }^{3}\int \limits_{0}^{t} d \tau  e^{- \lambda  \tau} = \rho_0
 [  e^{- \lambda t} + (1 - e^{- \lambda t} )  ] =  \rho_{0 }.
\end{equation*}
Obviously the sequence $ \{ \rho^{(k) } (t), k =0, \dots \}   $
is monotone increasing and bounded. Taking its limit and the
limit of equations \eqref{eq-4.20}, as  $  k \to \infty $, we
can easily show that	
\begin{equation*}
\rho (t)    = \lim \limits_{ k \to \infty }\rho^{(k) } (t)   = \rho_{0 }.
\end{equation*}
The transition to the limit under the integral sign is justified
by Lebesgue's theorem on  dominant convergence here and below.
For brevity we ignore sets of zero measure. 
On the other hand, it follows from \eqref{2.10}, \eqref{2.11} that   $   0  \leq     \varphi^{(k) }_{i} \leq  \rho_{0 } $ 
for all $  k \geq 0  $ and $ i=1,\dots, n  $.
Therefore
\begin{equation*}
\varphi(t)  = \{ \varphi_{i}(t) = \lim \limits_{ k \to \infty }
\varphi_{i}^{(k) }(t) ,  \quad
i=1,\dots, n  \}  
\end{equation*}
It follows from equations \eqref{eq-4.18}, \eqref{eq-4.19} that the function
   $ \varphi(t)     $  solves equation \eqref{eq-4.16}.  
Note that 
\begin{equation}\label{eq-4.21}	
\rho [   \varphi(t)  ]   = \rho_{0 } = const.
\end{equation}
in the notation of \eqref{eq-4.11}. Therefore the components  of
$      A(\varphi )  =  \{  A_i(\varphi ),     i =1, \dots, n \}  $ in
equation \eqref{eq-4.16} read (see \eqref{eq-4.12})
\begin{equation*}
A_i(\varphi )= Q_i(\varphi ) + g  \varphi_i(t) \rho_{0 }^{2},
\quad
i =1, \dots, n.
\end{equation*}
Hence, the equation \eqref{eq-4.16} can be written as
\begin{equation*}
\varphi(t)  e^{ \lambda t}  = f^{(0) } + \int \limits_{0}^{t} d \tau  e^{ \lambda \tau}  \{ Q  [\varphi(  \tau)]  + \lambda  \varphi(  \tau) \},
\end{equation*}
where  $ \lambda = g \rho_{0 }^{2} $. Then  we can prove by  differentiation that  $ \varphi(t)    $ solves the Cauchy problem
\begin{equation}\label{eq-4.22}	
\frac{d \varphi}{dt } = Q  ( \varphi ), \quad \varphi|_{t= 0} = f^{(0)}.
\end{equation}
The uniqueness of its solution follows from standard theorems for
autonomous ODEs with polynomial right hand side. Note that we did not use
any connection of coefficients $ \G_{ij}^{kl}  $ in equation \eqref{eq-4.3} with the set from \eqref{eq-4.4}.

Thus the following lemma is almost proved.
\begin{Lemma}\label{lem4.1}
We consider equations \eqref{eq-4.2}, \eqref{eq-4.3}. Then for any non-negative data
\begin{equation}\label{eq-4.23}	
f_{t=0} = f^{(0)} = \{ f_i \geq 0, \quad i=1,\dots, n   \}
\end{equation}
there exist a unique global in time solution $ f(t) $ of equations
\eqref{eq-4.2}, \eqref{eq-4.3}. The functions $  f_i(t) $	 satisfies inequalities
\begin{equation}\label{eq-4.24}	
f^{(0)}_{i } e^{ -\lambda t}  \leq  f_i(t) \leq \rho_{0}, \quad
  i=1,\dots, n; \quad  \rho_{0} = \rho ( f^{(0)} ) 
\end{equation}
in the notation of \eqref{eq-4.11}, \eqref{eq-4.13}.
\end{Lemma}
\begin{Proof} To finish the proof it is sufficient to note that the
problem \eqref{eq-4.22} for $   \varphi(t) $  coincides with the problem 
\eqref{eq-4.2}, \eqref{eq-4.3},	\eqref{eq-4.23} for $ f(t) $. Therefore
we just need to set $ f(t) = \varphi(t) $, where $\varphi(t)  $ was  already
constructed above. The inequalities  \eqref{eq-4.24} follow from
\eqref{eq-4.17}, \eqref{eq-4.11}.
This completes the proof.
\end{Proof}

It is clear that Lemma \ref{lem4.1} proves the first part of Theorem \ref{Th4.1}
(without the statement (b)) under much weaker conditions independent
of the set $ V $   from  \eqref{eq-4.4}  and specific properties of normal
discrete model of WKE. The role of these stronger conditions will be
 clear in the next section.

 \subsection{Existence of unique stationary solution}

 Our goal  in this section is to prove that the stationary equation
 \eqref{eq-4.2}, i.e. the equation
\begin{equation}\label{eq-4.25}	 
 Q(f) = 0
 \end{equation}
in the notation of \eqref{eq-4.3}, has a unique solution under assumptions 
of Theorem 4.1 and additional assumption  that 
\begin{equation}\label{eq-4.26}	 
\sum \limits_{i = 1 }^{n} f_i  = \rho, \quad
\sum \limits_{i = 1 }^{n} f_i |v_i|^2   =  E = \rho  T, \quad
\sum \limits_{i = 1 }^{n} f_i v_i= 0
 \end{equation}
in the notation of  \eqref{eq-4.4}. This fact follows from Theorem \ref{Th3.1} (c) from
Section 3.5, we just need to check that all conditions of that theorem are
satisfied. We note that the model \eqref{eq-4.2}--\eqref{eq-4.4} is
normal by  assumptions of Theorem \ref{Th4.1}. Moreover the function
$ p(x) = - 1/ x  $  satisfies  inequality  \eqref{eq-3.24}	 for
$ F(x_1,x_2; x_3,x_4) $ given  in \eqref{eq-4.1},  as it was shown in
\eqref{eq-3.33}. The function $ p(x) = - 1/ x  $ also satisfies
conditions (1) and (2) of Theorem \ref{Th3.1} from Section 3.5. Hence, we can use
 the part (c) of that theorem and conclude that any solution of \eqref{eq-4.25}  reads
\begin{equation}\label{eq-4.27}	 
f = \{ f_1, \dots, f_n \},\quad
f_i =  (   \alpha + \beta \cdot v_i + \gamma |v_i|^2 )^{-1},
\quad
 i=1,\dots, n,
 \end{equation} 
where $  \alpha  \in \R  $,  $ \gamma \in \R  $  and $  \beta \in \R^d $,
$  i=1,\dots, n $ are free parameters.  We
obtain these parameters from equations  \eqref{eq-4.26} and assumptions
of Theorem \ref{Th4.1} . Note that $ v_1= 0 $, $ v_i \neq v_j $  if $  i \neq j $,
and the set $ V  $ in \eqref{eq-4.4} is invariant (perhaps with change of
 numeration)  under transformation $ v_i \to (-v_i) $, 
 $ 1 \leq i \leqslant n $,
 as it follows from these assumptions. Hence,  $ \alpha \neq  0 $ in
\eqref{eq-4.27}  and we can change the notation of \eqref{eq-4.27} to
\begin{equation}\label{eq-4.28} 
 f_i = a (1 + b|v_i|^2 + \beta \cdot v_i)^{-1}, \quad
 1 \leq i \leq n.
 \end{equation}

  Then we apply the third condition from \eqref{eq-4.26}	 and obtain
\begin{gather} 
\nonumber
0 = a \sum \limits_{i = 1 }^{n} v_i   \psi(v_i) = \frac{a}{2}
\sum \limits_{i = 1 }^{n} v_i   [\psi(v_i) - \psi( - v_i)] =
\\
a \sum \limits_{i = 1 }^{n} v_i ( \beta \cdot v_i) [( 1 + b |v_i|^2)^2
-  (\beta \cdot v_i)^2]^{-1}, \quad
    \psi(v) = 1 + b |v|^2 + \beta \cdot v.
 \label{eq-4.29}   
 \end{gather}   
 We are interested in bounded non-negative functions $ f_i  $  in
\eqref{eq-4.28}.  Therefore 
 \begin{equation*} 
(1 + b \,|v_i|^2 )^2 > (\beta \cdot v_i)^2, \quad 1 \leq i \leq n.
\end{equation*} 
Then we multiply scalarly the equality  \eqref{eq-4.29} by the constant
vector $  \beta  \in \R^d  $ and obtain in the right hand side a sum
of non-negative numbers. Hence, $ \beta = 0  $, since the equality  
$ \beta \cdot v_i = 0  $ for all $ 1 \leq i \leq n $ is impossible for
any normal model  (see Definition 1).

Then  it follows from algebraic equations  \eqref{eq-4.26} that
\begin{equation}\label{eq-4.30} 
a= \frac{\rho}{S_0(b)}, \quad  
S_k(b) = \sum \limits_{i = 1 }^{n} |v_i|^{2k} ( 1 + b |v_i|^2 )^{-1},
 \quad k=0,1,\dots,
\end{equation} 
where $ b $ is maximal real root of algebraic equation

\begin{equation}\label{eq-4.31} 
T = \frac{S_1(b)}{S_0(b)}.
\end{equation}
The existence of such root follows from simple considerations. If we consider \eqref{eq-4.31} as the definition of function $ T(b) $, then we can
compute its derivative $ T'(b) $ and obtain
\begin{equation*}
T'(b) =  - \frac{1}{2} \sum \limits_{i,j  = 1 }^{n}  \left [
\frac{b (|v_i|^2 -|v_j|^{2}) }{ (1 +  b |v_i|^2)  (1 +  b |v_j|^2)}
\right ]^{2} >0.		
\end{equation*}
It is easy to see that $ T(b)  $ decreases monotonically from
$  T( - M^{-1 } ) = M= \max \{|v_i|^2 , \;  i=1,\dots, n  \}  $ to zero for
$  b  \to \infty $. Hence, the inverse function    $  b(T) $, satisfying
 \eqref{eq-4.31}, is uniquely defined for all   $0 < T < M $.
 The limiting value   $ b(M)=  - M^{-1 } $ means that the solution  \eqref{eq-4.28} becomes singular. This limit  is irrelevant for Theorem \ref{Th4.1} because we always have $ T_0 < M $ for initial data \eqref{eq-4.5}.

 The result of this section can be formulated as follows.
 \begin{Lemma}\label{lem4.2}
 We assume that conditions of Theorem \ref{Th4.1} are satisfied for equations
 \eqref{eq-4.2}, \eqref{eq-4.3} and the set  \eqref{eq-4.4}. Then the stationary equation \eqref{eq-4.25} has a unique non-negative solution
 \begin{equation}\label{eq-4.32} 
 f = f^{st} = \{ f^{st}_{1}, \dots, f^{st}_n\}, 
 \quad
 f^{st}_i = a (1 +  b |v_i|^2)^{-1}, \quad  i=1,\dots, n \}, 
 \end{equation}
 satisfying conditions \eqref{eq-4.26}  for any $ \rho > 0  $ and
 $   0 < T <  \max \{|v_i|^2 , \;  i=1,\dots, n  \}$. The parameters
 $ a $ and $ b $ are defined uniquely by algebraic equations
 \eqref{eq-4.30},   \eqref{eq-4.31}.
  \end{Lemma}
 The proof of Lemma \ref{lem4.2} is already given above.  We use the notation
 $ f^{st} $ in \eqref{eq-4.32} in order to distinguish the stationary
 solution from time-dependent solution $ f(t)  $
of equations \eqref{eq-4.2}, \eqref{eq-4.3}.

 In the next section we study some properties of the stationary
 solution  \eqref{eq-4.32}.

 \subsection{Properties of the stationary solution }
 It was shown in the beginning of the previous part that all conditions
 of Theorem \ref{Th3.1} are satisfied with function   $  p(x)  = - 1/x $  under assumption of Theorem \ref{Th4.1}. In particular, the inequality  \eqref{eq-3.24}
with $ F(x_1,x_2; x_3,x_4) $ from   \eqref{eq-4.1}   and 
$  p(x)  = - 1/x $ is given in  \eqref{eq-3.33}. Hence, we can apply the part (a) of Theorem \ref{Th3.1} from Section 3.5 and  equations     \eqref{eq-3.45},
 \eqref{eq-3.46}  from its proof. Obviously we can choose
 \begin{equation}\label{eq-4.33}  
 I(x) = -\log x, \quad   H[f(t)]   = -   \sum \limits_{i = 1 }^{n} \log 
 f_i(t)
 \end{equation} 
and conclude that 
 \begin{equation*}
 \frac{d}{dt} H[f(t)]  \leq 0
 \end{equation*}   
 on any positive solution  $ f(t) = \{ f_{1}(t)> 0, \dots, f_n(t)> 0  \} $  
of equations    \eqref{eq-4.2},  \eqref{eq-4.3}. One can check directly that
  \begin{equation}\label{eq-4.34}  
  \frac{d}{dt} H[f(t)]  = - \frac{1}{4} \sum \limits_{i,j,k,l = 1 }^{n}
  \G_{ij}^{kl} f_i f_j f_k f_l    (  f_i^{-1} +  f_j^{-1} 
   - f_k^{-1}- f_l    ^{-1} )^2 \leq 0
 \end{equation} 
 in accordance with equation \eqref{eq-3.33}. Thus, $ H(f) $ is the
 Lyapounov function for equations  \eqref{eq-4.2},  \eqref{eq-4.3}.
 
 Let us try to minimize $ H(f) $   in the domain $  \Omega \subset \R^n $
 such that
   \begin{equation}\label{eq-4.35} 
\Omega (\rho, T) =  \{   f = (f_1, \dots, f_n): \quad (1)\;\; f_i > 0,
\; i =1, \dots, n; \quad (2)  \;\;f \;\; \text{satisfies} \;  \eqref{eq-4.26} \},
\end{equation}  
where the set $ V = \{ v_1, \dots, v_n\} $ in \eqref{eq-4.4} satisfies
conditions of Theorem \ref{Th4.1}. It is straight-forward to see that the 
standard method of Lagrange multipliers shows that the point 
 $ f^{st}  \in \Omega $ from Lemma \ref{lem4.1} is a unique in $ \Omega $    point 
 of extremal. In fact  it is a point of minimum because
 \begin{equation*}
 \frac{\partial^2 H}{\partial  f_i \partial f_j}  =
 \delta_{ij }f_i^{-2} \geq 0,  \quad i,j = 1,\dots, n.  
 \end{equation*}  
 
 Hence, we can construct a modified Lyapounov function
    \begin{equation}\label{eq-4.36}
   U(f) = H(f) - H(f^{st}) \geq 0,
   \quad f \in \Omega (\rho, T) 
    \end{equation}  
 in the notation of Theorem 4.1. The point $ f =f^{st}  \in \Omega   $   
is the only point in $ \Omega  $ where $  U(f) = 0  $. The function 
 $ U(f) $    will be used in the next section for the proof 
 of Theorem \ref{Th4.1}.

    \subsection{Proof of convergence to equilibrium}

 We consider again the Cauchy problem   
  \begin{equation}\label{eq-4.37}
 \frac{d f}{dt}= Q( f ), \quad f|_{t=0} = f^{(0)}
     \end{equation} 
 in the notation of    \eqref{eq-4.2}--\eqref{eq-4.5} and assume that
 all conditions of Theorem \ref{Th4.1} are fulfilled. The unique solution
 of the problem \eqref{eq-4.37} was constructed in Section 4.2. This
 solution obviously satisfies conservation laws
 
    \begin{equation}\label{eq-4.38}
\sum \limits_{i = 1 }^{n} f_i(t) = \sum \limits_{i = 1 }^{n} f^{(0)}_{i} 
= \rho, \quad \sum \limits_{i = 1 }^{n} f_i(t) v_i =0,
\quad
 \sum \limits_{i = 1 }^{n} f_i(t)|v_i|^{2} = \sum \limits_{i = 1 }^{n} f_i^{(0)}|v_i|^{2} =    \rho T,
    \end{equation}   
in the notation of    \eqref{eq-4.4}. Moreover $  f_i(t) > 0 $  for all
$  1 \leq i \leq n $ and all   $  t> 0 $  because of lower estimates
in \eqref{eq-4.24}. Hence, for any $  t \geq 0 $ 
\begin{equation}\label{eq-4.39}
f(t) \in  \Omega =\Omega (\rho, T)
\end{equation}   
in the notation of  \eqref{eq-4.35}. Note that $ \Omega \subset \R^n $
is a bounded domain for any values of parameters $\rho  > 0  $, \;
$  0 < T < \max \{|v_i|^{2}, \quad i=1,\dots, n    \} $.
In particular, $  \Omega \subset B(\rho \sqrt{d}) $, where
$  B(\rho \sqrt{d})  $ is the ball of radius $ \rho \sqrt{d} $ 
centred at the origin.

We have constructed the Lyapounov function $ U(f) $
\eqref{eq-4.36} for equation \eqref{eq-4.37} such that  $ U(f) \geq 0 $
for any $ f \in \Omega $. It was also shown that there is a  unique point
$  f^{st}  \in \Omega $ such that $ U(f^{st}) =0 $.
Note that  the derivative
  \begin{equation}\label{eq-4.40}
\frac{d }{dt}   U[f(t)] = \frac{d }{dt}  H[f(t)] \leq 0
\end{equation} 
was computed in \eqref{eq-4.34} We can rewrite  \eqref{eq-4.40} as
 \begin{equation*}
\frac{d }{dt}   U[f] = grad_f \,U[f]\, \cdot Q[f] = - W[f],
\end{equation*}
where dot stands for the scalar product in $ \R^n $  and
  \begin{equation}\label{eq-4.41}
 W(f)= \frac{1}{4} \sum \limits_{i,j,k,l = 1 }^{n}
  \G_{ij}^{kl} f_i f_j f_k f_l    (  f_i^{-1} +  f_j^{-1} 
   - f_k^{-1}- f_l    ^{-1} )^2 .
\end{equation} 
It is clear that under conditions of Theorem \ref{Th4.1} the equation
$   W(f)=  0 $ has a unique solution $ f = f^{st} $ \eqref{eq-4.32}
 in  $  \Omega  $. This fact was actually used in the proof of Lemma  \ref{lem4.2}.
 Hence, similarly to $ U(f) $ the function $ W(f) $ has the following
 properties:
   \begin{equation}\label{eq-4.42}
 (a) W(f) > 0 \; \text{ if }  f \in \Omega, \, f \neq f^{st};
 \quad
 (b) W(f^{st}) = 0, 
\end{equation}   
where  $  f^{st} \in \Omega $  is given in \eqref{eq-4.32}.

We are almost prepared to prove that 
$$  \lim \limits_{t \to \infty}  f(t) =  f^{st}$$
on the basis of well-known facts from the theory of ODEs. To this goal
we begin with the following lemma.
\begin{Lemma} \label{lem4.3}
The above constructed solution  \eqref{eq-4.39}  satisfies for all $ t \geq 0 $ the inequality 
\begin{equation}\label{eq-4.43}
f_i(t) \geq \rho^{-(n-1) } \prod \limits_{j=1}^{n} f_j(0),
\quad
\rho =  \sum \limits_{j = 1 }^{n}  f_j(0),
 \quad i=1,\dots, n.
\end{equation}  
\end{Lemma}

\begin{Proof}
We consider the function $ H(f)  $ \eqref{eq-4.33}  and note that 
$ H[f(t)  ]  \leq  H[f(0)  ]  $ because of inequality  \eqref{eq-4.34}
or, equivalently 
$$ \prod \limits_{i=1}^{n} f_i(t)  \geq 
\prod \limits_{i=1}^{n} f_i(0).  $$
Since $ f_i(t) \leq \rho $ for any $ i=1,\dots, n $, we obtain a lower estimate for each component of $ f(t) $:
$$   f_i(t) \geq \rho^{-(n-1) } \prod \limits_{j=1}^{n} f_j(0),
\quad 
 i=1,\dots, n. $$
This completes the proof.
\end{Proof}

Hence, for any trajectory $ f(t) $ satisfying  \eqref{eq-4.39}, we can 
introduce a closed bounded domain $ \Omega_1  \subset \Omega $ such that
\begin{equation}\label{eq-4.44}
\Omega_1   = \{ f= (f_1, \dots, f_n) \in \Omega :
\quad
f_i \geq \rho^{-(n-1) }   \prod \limits_{j=1}^{n} f_j(0)   \}.
\end{equation}  
It follows from Lemma \ref{lem4.3} that  $ f(t) \in  \Omega_1   $ , $ t \geq 0 $,
where     $ f(t)  $ is the solution of the problem \eqref{eq-4.37}.
Note that  the definition of $ \Omega_1 $  depends on the initial
data  not only through sums \eqref{eq-4.38}, but also though the product of
components of $ f^{(0)} $.

We recall some known applications of Lyapounov functions (see e.g. the
textbook in ODEs \cite{TVS} ). With some abuse of notation, we consider a vector
ODE (like \eqref{eq-4.2}) 
\begin{equation}\label{eq-4.45}
\frac{df}{dt} = Q(f),
\quad
f \in \R^n, \quad  Q(f) \in \R^n.
\end{equation}  

It is assumed for simplicity that components  $  Q_i(f),
 \quad  1 \leq i \leq n $ are polynomials in components  of
 $ f = (f_1, \dots, f_n) $. Of course, these polynomials can
 differ from those shown in \eqref{eq-4.3}. Let $  D \subset \R^n  $
  be a closed bounded domain and there exist a solution $ f(t)  $ of  equation \eqref{eq-4.45}   such that  $ f(t) \subset D $  for all
  $ t \geq 0 $. The following theorem is a simple modification of
  Theorem 5.3 from \cite{TVS}. 
  In fact the first version of this theorem
  was proved by A.M. Lyaupunov in 1892 in his thesis \cite{Lyap}.
  
  \begin{Theorem}\label{Th4.2}
  It is assumed that equation \eqref{eq-4.45}  has a Lyapounov function $  U(f)$ such that 
  
  (a)  $  U(f) \geq 0 $   for all $  f \in D $, where $ D \subset \R^n $ is
  closed bounded domain;
  
  (b) $  W(f) = - grad_f U \cdot Q(f) \leq 0 $ for all $ f \in D $;
  
  (c) both functions   $  U(f) $     and   $W(f)  $ are continuous in $ D $;
  
  (d) there exists a unique vector (a point in $\R^n $)  $ f^{st}  \in D $
such that   
 \begin{equation}\label{eq-4.46} 
 U( f^{st} ) = W( f^{st}  ) = 0.
\end{equation}  
Then any solution $ f(t) $ of \eqref{eq-4.45} such that $ f(t) \in D $ for all $ t \geq 0 $  converges to $ f^{st}  $, i.e.
 \begin{equation}\label{eq-4.47} 
 \lim \limits_{t \to \infty} f(t) = f^{st}
\end{equation}  
  \end{Theorem}
  \begin{Proof}
  The proof is simply a repetition of the proof of Theorem 5.3 from \cite{TVS}.
  Therefore we just outline the scheme of the proof. The first step is to
 prove that $  U[ f(t) ]  \to 0 $, as $ t \to \infty $. Assuming the opposite
   we obtain that 
   $  U[ f(t) ]  \geq \alpha > 0, \,t \geq 0 $. Then we can prove by contradiction  that $ | f(t)  - f^{st}  | \geq \beta $ and
 $ W [ f(t) ]  \geq \gamma $  for all  $  t  \geq 0 $.  Hence we obtain that
\[
\frac{d U [ f(t) ] }{dt} \leq - \gamma \Rightarrow U[ f(t) ]   \leq U[ f(0) ]  - \gamma t, \quad t  \geq 0 .
\] 
This obviously contradicts to assumption that  $ f(t) \in D $  and 
therefore $ U[ f(t) ]  \geq 0 $  for all $  t  \geq 0 $. Hence,
$ U[ f(t) ]  \to 0 $,  as $  t \to \infty $. The second step is to prove 
the limiting equality   \eqref{eq-4.47}. Again we assume the opposite. Then there exist $ \varepsilon > 0  $  and a sequence
$  \{t_k, k \geq 1 \}    $ such that $ t_n \to \infty $, as
$ n \to \infty $, but 
 \begin{equation}\label{eq-4.48} 
|f(t_k)  - f^{st} | >  \varepsilon
\end{equation}   
for all $ k \geq 1 $. The sequences $ \{ f_k = f(t_k), 
\quad
 k=1,2,\dots \}   \subset   D $  is bounded and therefore contains a convergent subsequence, which converges to a point $ \bar{f} \in D $  (here we need the domain $ D $ to  be closed).  The equality
 $ \bar{f}  = f^{st} $ is impossible because  of inequalities
 \eqref{eq-4.48}. If we substitute  this convergent subsequence
 into the continuous Lyapounov function $ U(f) $, then the
 corresponding sequence converges to $ U(\bar{f}) \neq 0 $. Hence,
 we obtain a contradiction and this completes the proof.
  \end{Proof}

\textbf{ The end of the proof of Theorem \ref{Th4.1}.} It remains to apply Theorem \ref{Th4.2} to the case of equations \eqref{eq-4.2},  \eqref{eq-4.3}  with initial data \eqref{eq-4.5}. We consider the
 solution  $ f \in \Omega_1 $  constructed in  Lemma \ref{lem2.1}.
 It is clear that all conditions of Theorem \ref{Th4.1} are fulfilled. Indeed we choose the domain $ D = \Omega_1 $
 in the notation of \eqref{eq-4.44} and 
the functions $ U(f) $  and $ W(f) $ in the notation of  \eqref{eq-4.27} and \eqref{eq-4.41}, respectively. The stationary solution  $ f^{st}   \in  \Omega_1 $
 was constructed in Lemma 4.2 in explicit form \eqref{eq-4.32}
 under assumptions of Theorem 4.1.
 Equalities \eqref{eq-4.46}	 follow from formulas for $ U(f) $
\eqref{eq-4.36} and  $ W(f) $ \eqref{eq-4.41}  under the same
assumption. The uniqueness in $  \Omega $ (and therefore in 
$  \Omega_1 \subset  \Omega $) of the root $ f = f^{st} $ of
equations 
$ U(f)= W(f) = 0 $ was proved above, see \eqref{eq-4.47} and
comments after \eqref{eq-4.36}. Hence, the assumptions of
Theorem 4.1 allow us to apply Theorem 4.2 and prove the limiting
equality \eqref{eq-4.7}. This completes the proof of Theorem 4.1.

	\subsection{Conclusions}
 
A large class of nonlinear kinetic equations of the Boltzmann type was considered in Sections 2--4 from a unified point of view.  This class includes, in particular, such well-known equations  as  
(a) the classical Boltzmann  equation, (b) the quantum Nordheim--Uehling--Uhlenbeck equation, 
(c) the wave kinetic equation used in the theory of weak turbulence. 

It was shown that all these equations can naturally considered as different forms of the general Boltzmann-type equation introduced in Section 3. The general properties (conservation laws and monotone functionals) of that equation are  also studied there. By analogy with  discrete velocity models of the Boltzmann equation  the class of discrete models of the general kinetic equation was introduced and the properties of the models were studied.  

The long-time behaviour of solutions to discrete models of WKE was investigated in detail in Section 5. First we have proved the existence of unique global in time solution of the corresponding set of  ODEs  for any non-negative initial conditions. The Lyapunov function was constructed for any positive solution of the model and then used for the proof of convergence to equilibrium at the end of Section 5. This result is proved for so-called normal models which do not have any spurious conservation law. Perhaps, similar results can be proved also for discrete models of NUU-equation for fermions, but the case of WKE looks more interesting for some reasons. 

The matter is that it is natural to expect that the time-evolution of solutions to normal discrete kinetic models imitate, to some extent,  the behaviour of corresponding solutions to kinetic equations. In principle, we can approximate with any given accuracy the kinetic equation by a sequence of discrete models with sufficiently large number of discrete phase points, as it is shown in Appendix B. These arguments work very well in the case of the Boltzmann equation, for which the discrete models predict that a solution $f(v,t)$,  $ v \in R^d $, with finite moments 
up to the second order tends, as $t \to \infty $ to a Maxwellian distribution of the form 
$ M(v) = a \exp (- b |v|^2) $, 
under some irrelevant extra conditions. Positive parameters $  a $  and  $ b $   are determined by conservation laws. This prediction is absolutely correct for the Boltzmann equation.  On the contrary, the attractor for solutions to discrete models of WKE has the form  $f^{st} (v) = a (1 + b   |v|^2)^{-1}$.
 Obviously this function is not integrable in $ R^d $,
  $ d \leq 2 $  for any positive $ a $ and  $ b $. Therefore it cannot be an attractor for integrable solutions of WKE. Therefore a straightforward prediction of similar long-time behaviour is impossible in that case. At the same time the information about long-time behaviour of solutions to discrete models of WKE still can be useful.  We hope to come back to  
this question in subsequent publications.

\vspace{10mm}
\textbf{Acknowledgement} This work is supported by the Ministry of
Science and Higher Education of the Russian Federation (megagrant
agreement No 075-15-22-1115). 
I thank  S.B. Kuksin for important discussions. I am also grateful to I.F. Potapenko for help in preparation of the manuscript.


\subsection*{Appendix A. Construction  of normal discrete kinetic models}

\addcontentsline{toc}{subsection} {\textbf{Appendix A. Construction and classification of normal discrete kinetic models}}


We note that the definition of normal 	models from Section 3 includes only
conditions (a), (b), (c) that do not depended on function $F(x_1, x_2; x_3, x_4)$.
Therefore we just need to use some results obtained for
discrete velocity models of
the \BE  in \cite{Ved}, \cite{VO}, \cite{BC}, \cite{BV}.
 Moreover the condition (a) does not
depend on equations \eqref{eq-4.2}, it depends only on the set
 $  V = \{v_1, \dots, v_n \}
\subset \R^d  $  from \eqref{eq-4.1}. We denote

\begin{equation}	\tag(A1) \label{A1}
v_i = (v_i^1, \dots, v_i^d ), \quad d \geq 2, \quad 1 \leq i \leq n,
\end{equation}		
and introduce $ n$-dimensional vectors
\begin{equation}	\tag(A2) \label{A2}
\varphi_1=(1,1,\dots,1) ; \quad
\varphi_{\alpha +1} = (v_1^\alpha,  \dots, v_n^\alpha);
\;\alpha  =1, \dots, d, \quad
\varphi_{\alpha +2} = (|v_1|^2, \dots,|v_n|^2 ).
\end{equation}	
It is easy  to verify that the condition (a) means that vectors
 $ \{ \varphi_i\in \R^n, i=1 ,\dots, d+2\} $ are linearly independent.
We also note that the simplest Broadwell model with four point
 is not normal because all the points
lie on the circle. Therefore $ n \geq 6 $ for normal 	models.

Conditions (a) and (b)  of normality depend on the set of coefficients
$ \G_{ij}^{kl} > 0 $,  $1 \leq i,j,k,l \leq n  $.  It was noted in Section 3.4 that four points
 $\{ v_i, v_j; v_k,v_l\} $ form a rectangle in $ \R^d $
 if $ \G_{ij}^{kl} > 0 $. If we imagine all such rectangles connecting
all the $n$ (distinct) points $ v_i, \; 1 \leq i \leq n $, of the set $ V $,
 then the condition (b) means that each point of $ V $ belongs to at least
one such rectangle. Otherwise the model is not normal and we need
to drop isolated points. It is assumed below that conditions (a) and (b) are fulfilled.

Then it remains to discuss the condition (c). If
$ \G_{ij}^{kl} > 0 $ in equations \eqref{eq-4.2}, then we can say that the reaction
 $\{ v_i, v_j; v_k,v_l\} $ is possible.
It is clear  that all integers $i,j; k,l$   are distinct.
Then we can introduce the vector of reaction \cite{VO}

\begin{equation}	\tag(A3) \label{A3}
\theta_{ij}^{kl} = (\dots\underbrace{1}_i
\dots \underbrace{1}_j \dots \underbrace{-1}_i  \dots \underbrace{-1}_l) \in \R^n,
 \quad
\G_{ij}^{kl} > 0,
\end{equation}	
where dots stand for zeroes. In other words, the rule is to put $ (+1) $ on
positions $i,j$ and to put $(-1)$  on positions $k,l$. The next step
shows the convenience of this notation. Indeed the equation \eqref{eq-4.6} can
be written as the usual scalar product of two vectors from $ \R^n $

\begin{equation}	\tag(A4) \label{A4}
\theta_{ij}^{kl} \cdot h(v) = 0, \quad \G_{ij}^{kl} > 0,
\end{equation}	
where
$ h  = h_1, \dots,h_n  \in  \R^n$, \quad $h_i=h(v_i), \quad   1 \leq i \leq n$.
We can reformulate   the condition (c) in the following way: if the equality
{(A4)} is valid for any such indices $ (i,j;k,l) $  that  $  \G_{ij}^{kl} > 0 $,
then $ h(v) $ is a linear combination of vectors $\varphi_1(v), \dots, \varphi_{d+2}(v) $
from {(A2)}.

Let us consider all equalities  {(A4)}, where $\G_{ij}^{kl} > 0 $ . These equalities
are not linearly independent because, for example,
$\theta_{ij}^{kl}  = - \theta^{ij}_{kl}  $  by construction of vectors  {(A3)}. The
set {(A4)} of such equalities (or,  equivalently, equations for  $h(v)  \in  \R^n $ )
is obviously  equivalent to its subset

\begin{equation}	\tag(A5) \label{A5}
\theta_{i_\beta j_\beta }^{k_\beta  l_\beta } \cdot h(v) = 0,
\quad
 \beta =1, \dots,  p,
\end{equation}	
where vectors

\begin{equation}	\tag(A6) \label{A6}
\{\theta _{i_\beta j_\beta }^{k_\beta  l_\beta }, \quad  \beta =1, \dots,  p \}
\end{equation}	
are linearly independent and $p$ is the maximal number of such vectors of the form
({A3}). It is easy to see that
\begin{equation}	\tag(A7) \label{A7}
p \leq  p_{max}, \quad p_{max}  = n -(d+2),
\end{equation}	
because all vectors ({A3) are by construction orthogonal to $(d+2)$
linearly independent vectors $ \varphi_1, \dots, \varphi_{d+2} $ from ({A2}).
We introduce two orthogonal subspaces of $ \R^n $

\begin{equation}	
\tag(A8) \label{A8}
\Phi =Span\,\{\varphi_\alpha,\; \;\alpha=1, \dots, d+2\},
\quad
\Theta= Span\,\{ \theta _{i_\beta j_\beta }^{k_\beta  l_\beta }, \quad  \beta =1, \dots,  p \}.
\end{equation}	
Then if    $ p < p_{max} $ we can represent   as an orthogonal sum of three spaces
\[ \R^n = \Phi  \oplus \Theta  \oplus H_\perp \]
where $ H_\perp  $ contains all solutions of equations ({A5}) that are
orthogonal to $  \Phi    $.   It is clear that
 $    dim H_\perp = n -( p + d+2) $.  Hence,
 there is only one  possible
case $ p    =  p_{max} $, when the subspace $ H_\perp  $ is empty. Thus, the following fact
is proved (see also  the original proof in \cite{VO}).

\begin{Lemma} 
It is assumed that conditions (a) and (b) of Definition 1  are fulfilled for given model
\eqref{eq-4.1}, \eqref{eq-4.2}.   Then the condition (c) is also fulfilled if and only if the set
of all its vectors of reaction ({A3}) contains  $p = n (d+2) $ linearly independent vectors.
\end{Lemma}

The following consideration shows an inductive way of constructing normal model. We
fix the dimension $ d \geq 2  $ of vectors  $v_i \in V, \quad i=1, \dots, n $ and assume
that the model \eqref {eq-4.1}, \eqref{eq-4.2} is normal.  Then $ dim \Theta = n - (d+2)  $
 in the notation of ({A8}).  We assume in addition that there are three points in $V$,
say, $  v_{n-2},  v_{n-1},  v_{n} $ such that

\begin{equation}	\tag(A9) \label{A9}
( v_{n} - v_{n-1} ) \cdot (  v_{n} - v_{n-2} ) = 0,
\quad
 v_{n+1}   = v_{n-1} + v_{n-2}     - v_{n} \nsubseteq \in  V.
\end{equation}	
Then we can add a new point      $  v_{n+1} $ to the set  $ V $  and a new vector of reaction

\begin{equation}	\tag(A10) \label{A10}
\theta_{n-1 \; n-2}^{n \; n+1} \{ 0, \dots,1,1,-1,-1 \}  \in \R^{n+1},
\end{equation}	
where dots stand for zeros. Indeed the new reaction satisfies conservation laws

\begin{equation*}
v_{n-1}  +   v_{n-2} =  v_{n} +  v_{n+1},
\quad
|v_{n-1}|^2  +  | v_{n-2}|^2 = |  v_{n} |^2 + | v_{n+1}|^2
\end{equation*}	
and therefore the inequality $ \G_{n-1 \; n-2}^{n \; n+1} >  0 $ is allows
for extended model. The linear independence of new vector ({A10})
with vectors from  $ \Theta $ ({A8}) extended to $ \R^{n+1}  $ is obvious
because only vector ({A10}) has a non-zero component on the $  (n+1)^{th} $
position. Hence, we obtain a new normal model, where the set
 $ V \subset \R^d $
and  the vector  $ f(t)  \in \R^n $  from \eqref{eq-4.1} are replaced respectively by

\begin{equation*}
\hat{V} = \{  v_1, \dots, v_{n+1}   \}   \subset \R^d,
 \quad
\hat{f}(t) = \{   \hat{f}_1(t), \dots,   \hat{f}_{n+1} (t)   \}  \in \R^{n+1}.
\end{equation*}
The changes in equations \eqref{eq-4.2} are obvious. In particular the equation
for $   \hat{f}_{n+1} (t)   $ reads

\begin{equation}	\tag(A11) \label{A11}
\frac{d  \hat{f}_{n+1} (t) }{dt}  = \G_{n-1 \; n-2}^{n \; n+1}
\, F( f _{n-1}, f_{n-2}; f_{n} , f_{n+1} ).
\end{equation}

Of course, we need an initial normal model to begin  the process of successive extensions.
If we want to get regular lattices, then it is convenient to begin with $d$-dimensional
Broadweell model with

\begin{equation}	\tag(A12) \label{A12}
V= \{ e_1, -e_1, e_2, -e_2, \dots, e_d, -e_d \}  \subset \R^d,
\end{equation}	
where  $e_i, \; i= 1, \dots, d $, are unit vectors directed along Cartesian coordinates axes.
It is easy to see that the set $V$ can have at most   $ p =d-1  $ linear  independent vectors
({A3}).
 Note that the discrete model with the set $ V $ from
({A12}) cannot be normal by definition because
all points of $V$ lie on the unit sphere $S^{d-1}  $. Then we add two points to $V$,
namely,  $   v_{2 d + 1} = 0 $ and $   v_{2d+2} = e_1 +  e_2 $ . Then we obtain
a  normal model with

\begin{equation*}
V= \{  v_1, \dots, v_{n}   \} , \quad
n=2(d+1),
\end{equation*}
where

\begin{equation*}
 v_{2k+1} = e_k, \quad  v_{2k+2} =  - e_k, \quad
k=0, \dots, d-1;
\quad
 v_{2d+1} = 0, \quad
v_{2d+2} = e_1+e_2.
\end{equation*}
It is easy to check that this model is normal. Indeed it has $(d-1)  $ linearly  independent
vectors ({A3}) of the form  $   \theta_{1\; 2}^{2k+1\; k+2}  $, \,  $ k=1, \dots, d-1 $,
and also $   \theta_{1\; 3}^{n-1\; n}   $  because
\[ v_1=e_1, v_3=e_2,v_{n-1}=0, v_n= e_1+e_3 \]
and  $ e_1 \cdot e_2 = 0 $ by construction. 
Hence, we obtain  $  p = d  $ linear independent
 vectors ({A3}) for
 of  $  n = 2d + 2 $ points. Then $   p= n - (d+2) $ and therefore it follows from
Lemma 4.4 that  this model is normal.
Thus we can use an inductive way of  enlarging this model by adding to it all vectors
(one by one) of the form $ v  =   v_i +v_j, \; i \neq j$,
and so on.

\subsection*{Appendix B. On approximation of Boltzmann-type equations \\by discrete kinetic models}
\addcontentsline{toc}{subsection} {\textbf{Appendix B. On approximation of Boltzmann-type equations by discrete kinetic models}}

	We consider below the Boltzmann-type equation \eqref{eq-3.1}
	for the distribution function $ f(v,t) $, $ v \in \R^d $,
	where $ K[f] $ is written in the form \eqref{eq-3.12}. Omitting
	irrelevant constant factor in \eqref{eq-3.12}, we obtain 	
\begin{equation} \tag {B1}\label{B1}
		f_t(v,t) = K[f](v)
=\! \!\int \limits_{\R^d \times S^{d-1}} \! \! dw\, d\omega\, |u|^{d-2} R(u, u')  \;
G(v, w; v', w'),
	\end{equation}	
where $  R(u, u') = R(u', u) $,

\begin{equation} \tag {B2}\label{B2}
G(v, w; v', w')= F[f(v), f(w); f(v'), f(w')],
\end{equation}	
\begin{equation} \tag {B3}\label{B3}
F(x_1, x_2;  x_3, x_4) = F(x_2,x_1 ; x_3,x_4) = F(x_1,x_2 ; x_4,x_3) = -F(x_3,x_4 ; x_1,x_2),
\end{equation}	
\begin{equation} \tag {B4}\label{B4}
\omega \in S^2,
\quad
u = v - w,\quad v' = (v + w +u') / 2,\quad 
u'= |u| \omega, \quad  w' = (v + w -u') / 2. 	
\end{equation}

Suppose that we want to approximate the integral 
 $ K[f](v) $ by a quadrature formula on some discrete
 lattice in the $ v $-space. To this end we introduce a regular grid in $ \R^d $
 $   \{ v_i \in \R^d , \quad i=1,\dots,n\} $.
 Let $\tilde{f}_{i}(t) $ be an approximation of $ f(v_i,t) $.
 Then a discrete version of equation \eqref{B1} reads
\begin{equation*} 
 \frac{d \tilde{f}_{i}}{dt} = \tilde{K}_i(\tilde{f}),
 \quad
 \tilde{f}= (\tilde{f}_1, \dots, \tilde{f}_n),
 \quad
 i=1, \dots, n,
\end{equation*}
where $  \tilde{K}_i(\tilde{f})   $ denotes a quadrature formula 
for $ K[f](v_i) $.  In principle, we can use any such formula,
but there are obvious advantages associated with approximations
of the form
\begin{equation*}
K_i(f)= \sum_{j,k,l=1}^{n}
	 \G_{ij}^{kl}  F(f_i,f_j ; f_k,f_l).
\end{equation*}
The tildes are omitted here and below. It is assumed that the constant
coefficients $ \G_{ij}^{kl} \geq 0 $, moreover $ \G_{ij}^{kl} > 0 $ only if
\begin{equation*}
v_i + v_j = v_k + v_l,\quad |v_i|^2 + |v_j|^2 = |v_k|^2 + |v_l|^2.
\end{equation*}  
Then we obtain the general discrete kinetic model of  equation
\eqref{B1} 
  \begin{equation} \tag{B5} \label{B5}
\frac{df_i}{dt} = \sum_{j,k,l=1}^{n}\;
	 \G_{ij}^{kl} F(f_i,f_j ; f_k,f_l),
	\quad
	\quad
	i=1, \dots, n.
\end{equation}  

Such models are discussed above in Sections 3.4--4.5 under
assumptions that
\[\G_{ij}^{kl} = \G_{ji}^{kl} = \G^{kl}_{ji}.\]
It is shown  there that solutions of the system \eqref{B5}	
of ODEs inherit the properties of the solutions of the kinetic
equation \eqref{B1}, namely, the conservation laws and $ H $-theorem
(if $ H $-theorem is valid for equation  \eqref{B1}).

	Our goal is to explain some methods of construction of such
	approximations. Our presentation can be considered as 
	a generalization of papers \cite{BPS}, \cite{PSB}, 
	devoted to similar problems for the classical \BE.

	Let $ (k_1 h, \dots, k_d h) $ be our discrete set of points
	in $ \R^d $, where  $ h $ is any positive number (mesh step)
and  $ (k_1, \dots, k_d ) $ are integer numbers. We identify this grid with $ \mathbb{Z}^d $ and call its points "integer points". We also use below the term "even" points if all $ n_i $
are even numbers.

	A natural first step is to use the simplest rectangular formula
  \begin{equation} \tag{B6} \label{B6}	
	 K[f](v_i) \approx (2h)^d \sum_{v_j} \int \limits_{S^{d-1}} d \omega  R(u_{ij}, |u_{ij}| \omega) G(v_i, v_j; v_i',v_j')
\end{equation} 	
in the notation of Eqs. \eqref{B1}--\eqref{B4}. Here $ v_i $
is a given point of the lattice and the sum is taken over all
such points $ v_j $ that  $ u_{ij}= v_i - v_j $ (this choice of
$ v_j $ will be explained below).

Then the next step is to choose an approximation of the 
integrals over unit sphere $ S^{d-1} $
  \begin{equation} \tag{B7} \label{B7}
I(v_i,v_j) =
 \int \limits_{S^{d-1}} d \omega  R(u_{ij}, |u_{ij}| \omega) G(v_i, v_j; v_i',v_j').
\end{equation} 	
If two integer vectors  $ v_i  $,  $ v_j $ are chosen in the 
above explained way, then $ u_{ij}= v_i - v_j $  is an even vector and
$ U_{ij}  = (v_i+ v_j)/2 $ is  an integer vector. We introduce
 an abbreviated notation
  \begin{equation} \tag{B8} \label{B8}
\varphi(|u_{ij} |^2/4,\omega) = | S^{d-1} |
R(u_{ij}, |u_{ij}| \omega) G(v_i, v_j; v_i',v_j'),
\end{equation} 	
where $ v_i'  $,  $ v_j' $  are given in \eqref{B4},
$| S^{d-1} |  $ denotes the area  of unit sphere in $ \R^d $.
Then the integral \eqref{B7} reads
  \begin{equation} \tag{B9} \label{B9}
I(v_i, v_j) = \frac{1}{| S^{d-1} | }    \int \limits_{S^{d-1}} d \omega \varphi(|u_{ij} |^2/4, \omega),
\end{equation} 	
where $ v_i \in h \mathbb{Z}^d $,  $ v_j \in  \mathbb{Z}^d $,
$  u_{ij} =  v_i -v_j \in 2 h  \mathbb{Z}^d $ by construction.
It is implicitly assumed that the function 
$ \varphi(|u_{ij} |^2/4,\omega) $ is known only in such points
$ \omega_k $, where $ v_i'  $ and   $ v_j' $  in \eqref{B7} 
are integer vectors. Let us assume that
 \begin{equation} \tag{B10} \label{B10}
u_{ij} = 2h (n_1, \dots, n_d),
\quad
|u_{ij} |^2 = 4 h^2 m,
\quad
 m = \sum \limits_{l=1}^{d}  n_l^2.
\end{equation} 
Then it is clear from formula \eqref{B4} for $ v' $ and
$ w' $ that  $ v_i' $ and $ v_j' $ in
\eqref{B8} are integer vectors if and only if
 \begin{equation*}
u_{ij} = v_i' - v_j'= 2h (x_1, \dots, x_d) \in 2 h \mathbb{Z}^d,
\end{equation*}
where
 \begin{equation} \tag{B11} \label{B11}
x_1^2 + \dots + x_d^2 = m
\end{equation} 
in the notation of 	\eqref{B10}. In our case a natural idea of
approximation of the average value \eqref{B9} of the function
$  \varphi(h^2 m, \omega) $ over the unit sphere $ S^{d-1} $ is
to replace the integral  \eqref{B9}  by the average value over 
"integer" (in the above discussed sense) points of the unit sphere.
In other words, we assume that for large $ m $
 \begin{equation} \tag{B12} \label{B12}	
 \begin{split}
	 \frac{1}{| S^{d-1} | }    \int \limits_{S^{d-1}} d \omega \varphi(h^2 m, \omega) \approx \frac{1}{r_d(m) }
	 \sum \limits_{x \in V_d (m)} \varphi(h^2 m, \frac{x}{\sqrt{m}}),
	\\
V_d (m) = \{ x=(x_1,\dots,x_d) \in  \mathbb{Z}^d,
\; x^2=m \},	
\end{split} 
\end{equation} 	
where $ r_d(m)$ is the number of integer solutions of equation
\eqref{B11}, i.e. the number of elements in $ V_d(m) $.
The first argument $ h^2 m $  of $ \varphi(h^2 m, \omega) $
is not important. We normally consider such limit in the general 
approximate	formula  \eqref{B6} that $ h \to 0 $, $ m \to \infty $, $ h^2 m = const $.

Thus, the second approximate formula 	\eqref{B12} is closely
connected with classical number-theoretical problem of finding
integer solutions of equation \eqref{B11}. There is a large
literature on this problem, see e.g. books \cite{Lin}, \cite{Gro}.
It is intuitively clear that approximate  equality 
\eqref{B12} becomes exact for continuous function of 
$ \omega \in S^{d-1} $	 in the limit $ m \to \infty $ provided
(1)	\; the number $ r(m) $ of solutions of \eqref{B11} tends to infinity as  $ m \to \infty $, and \; (2) \; these solutions tend to be 
equidistributed on the sphere. This is true for
$ d \geq 4 $,   see  \cite{Lin}	for details. Then situation, however, 
is less simple in more practically interesting cases 
	$ d=2 $	  and  $ d=3 $.

We briefly discuss these two cases below. All details can be found in
papers	
	\cite{BPS}, \cite{PSB} for  $ d=3 $	 and  \cite{FKW} for $ d=2 $.
	We begin with the case $ d=3 $. The	first difficulty is that equation   
\eqref{B11} for		 $ d=3 $ 	does not have integer solutions if 
$ m  = 4^{a}(8k + 7) $, where $ a $ and $k$ are integers. Fortunately
this difficulty is irrelevant in our problem, since it is known from
equality for $m$ in \eqref{B10} that there exists at least one integer solutions of 
	\eqref{B11}.
	
	Still there are "bad" sequences of the form $ m_{a}  = 4^{a}m_0 $,
	$ a\to \infty $, because $ r( m_{a} ) = r(m_0) $ for any
	 $  a=0,1,\dots  $. It follows from elementary observation that
$ x_1 $, $x_2$ and $x_3$ in \eqref{B11} are even if and only if 
$ m=0   $	(mod4). If $  m = 4^{a} m_1$, where $ m_1\neq 0 $ (mod4) and
$ m_1 \neq 7 $ (mod8), then   $ r_3(m) \to \infty  $, as  $m_{1}  \to \infty$.
	 	Of course, this is not enough for rigorous justification of approximate
	 	formula 	\eqref{B12}. The more difficult problem of equidistribution of solutions of 	\eqref{B11} on the sphere of radius $ \sqrt{m}$  in
	 	$\R^3  $ was proved in 	\cite{BPS}, \cite{PSB}  on the basis of rather
	 	complicated number theoretical results by Iwaniec \cite{Iwa}  (see also
	 	\cite{GF},\cite{Duk} and \cite{Sar}).
Without going in details we cite here one of results of  \cite{PSB} (see
Corollary 4 on p. 1873 there).

\begin{Prop}  \label{PropB1}		 
Let  $ \varphi (\omega)$	be a continuous function on $ S^2 $.  Then
 \begin{equation} \tag{B13} \label{B13}
\frac{1}{r_3(m)} \sum \limits_{x \in V_3(m)}\;\varphi \left(
\frac{x}{\sqrt{m}}\right) 
\xrightarrow[m \to \infty]
\; \frac{1}{4 \pi} \int \limits_{S^2}
d \omega \varphi (\omega)
\end{equation}
for every $ m=1,2,3,5,6 $ (mod8). 
\end{Prop}

Then we substitute the left 
hand side of approximate formula \eqref{B12} in the notation of
\eqref{B8} into approximate equality  \eqref{B6} and obtain for
$ d \geq 3  $
 \begin{equation} \tag{B14} \label{B14}
K[f](v_i) \approx K_h[f](v_i)=
(2h)^d \sum \limits_{v_j,v_k,v_l \in h \mathbb{Z}^d} \G_{ij}^{kl}\; F(f_i, f_j; f_k,f_l),
\end{equation}
where $ f_i = f(v_i) $, $ v_i \in  h \mathbb{Z} $,
 \begin{equation} \tag{B15} \label{B15}
 \G_{ij}^{kl} = \tilde{R}(v_i - v_i,  v_k-v_l)\frac{|S^{d-1}|}{r_d (|v_i-v_j|^2)
 /4h^2} \delta[  (v_i-v_j)^2 - (v_k-v_l)^2    ]\;
 \delta[ v_i+v_i -v_k-v_l ].
\end{equation}
Here $ \delta $-function of integer vectors and squares of such vectors denote simply corresponding Kronecker symbols. The kernel 
$ \tilde{R} (u,u') $ is defined by equality
\[
\tilde{R} (u,u') =
\begin{cases}
\tilde{R} (u,u'), & \text{if \;$u, u' \in 2h \mathbb{Z}^d $;} \\
0, & \text{otherwise.}
\end{cases}
\]
The convergence
 \begin{equation} \tag{B16} \label{B16}
K_h[f](v) \xrightarrow[h \to 0] \; K[f](v)
\end{equation}
is proved in \cite{PSB} for the classical Boltzmann equation with 
$ d=3 $ and for non-negative continuous  functions such that (see
e.g. \cite{Car})
 \begin{equation} \tag{B17} \label{B17}
 \parallel  f\parallel = \sup \frac{f(v)}{(1+ |v|^2)^d} < \infty.
\end{equation} 
This is the result of Theorem 11 from	\cite{PSB}. It can be easily generalized to the case $ d \geq 3 $ with $  \parallel  f\parallel  $
from \eqref{B17}	 and to the whole class of Boltzmann-type operators $ K[f](v) $   with function $  F(f_i(x_1),f_j(x_2) ; f_k(x_3), f_l(x_4))$
satisfying inequality

\begin{equation} \tag{B18} \label{B18}
 | F(f_i(x_1),f_j(x_2) ; f_k(x_3), f_l(x_4))|  \leq C(R) ( x_1 x_2 + x_3 x_4 )
\end{equation}
for all $ 0 \leq x_i \leq R $, $ 1 \leq i \leq 4 $, and for any
$ R > 0 $. We do not discuss here various estimates of the rate of convergence of quadrature formula \eqref{B14} for
$ h \to  0 $, obtained for the \BE in \cite{PSB}.

Instead we give some comments on the  difficult case $ d=2$ . In fact
 the quadrature formula \eqref{B14}  was also proved for the \BE under
 some smoothness assumptions on the functions under the integral sign
 in \eqref{B1}. Roughly speaking, it was proved in \cite{FKW} that
 lattice points  on circles are equidistributed on \underline{average} in the sense that  the exponential sums
 \[
 S(m,k) = \sum \limits_{|u'|^2 = m }\; e^{ik \theta_{u'}},
 \]
where $ \theta_{u'} \in [0, 2 \pi] $  is the angular coordinate of
$ {u'} \in \mathbb{Z}^2 $, converge to zero when $ m $ goes to
infinity. This is a key point in the proof in \cite{FKW} of the
quadrature formula \eqref{B14} for the plane case $ d=2 $.

 There is also another way of approximation of the Boltzmann-type 
 integrals \eqref{B1} by infinite sums. The idea is to use the
 Carleman form of the integral (see formula \eqref{eq-3.14} from
 Section 3.1) with inner integral  over a plane. The discrete models
 of the \BE based on this idea are constructed and discussed  in
 \cite{IP}. Independently, the convergence of certain infinite
 sums to the integral $ K[f] $ for WKE, written in similar form, was discussed in \cite{DymK}. For the sake of brevity, we do not discuss here the problem
 of approximation  of infinite sums by finite sums.
 It is always possible to do for functions $ f $ and $ F $, satisfying conditions 
 (B17) and (B18), respectively.


\end{document}